\documentclass[useAMS]{mn2e}
\usepackage{times}
\input{epsf}
\usepackage{graphicx}
\usepackage{lscape}

\def\apjl{ApJL}

\def\la{\mathrel{\hbox{\rlap{\hbox{\lower4pt\hbox{$\sim$}}}{\raise2pt\hbox{$<$}}}}}
\def\ga{\mathrel{\hbox{\rlap{\hbox{\lower4pt\hbox{$\sim$}}}{\raise2pt\hbox{$>$}}}}}

\title[The Ultraluminous State]
{The Ultraluminous State.}

\author[J. C. Gladstone, T. P. Roberts \& C. Done]
{Jeanette C. Gladstone$^{*}$, Timothy P. Roberts, Chris Done\\
$^1$Department of Physics, University of Durham, South Road, Durham DH1 3LE, UK\\
$^{*}$j.c.gladstone@durham.ac.uk}

\date{Accepted by MNRAS}
\pagerange{\pageref{firstpage}--\pageref{lastpage}}
\pubyear{2009}

\begin{document}
\label{firstpage}
\maketitle

\begin{abstract} 

We revisit the question of the nature of ultraluminous X-ray sources
(ULXs) through a detailed investigation of their spectral shape, using
the highest quality X-ray data available in the {\it XMM-Newton}
public archives ($\ga 10,000$ counts in their EPIC spectrum).  We
confirm that simple spectral models commonly used for the analysis and
interpretation of ULXs (power-law continuum and multi-colour disc
blackbody models) are inadequate in the face of such high quality
data. Instead we find two near ubiquitous features in the spectrum: a
soft excess and a roll-over in the spectrum at energies above
3~keV. We investigate a range of more physical models to describe
these data. Slim discs which include radiation trapping (approximated
by a $p$-free disc model) do not adequately fit the data, and several
objects give unphysically high disc temperatures ($kT_{\rm
in}>3$~keV). Instead, disc plus Comptonised corona models fit the data
well, but the derived corona is cool, and optically thick ($\tau \sim
5 - 30$). This is unlike the $\tau\sim 1$ coronae seen in Galactic
binaries, ruling out models where ULXs are powered by sub-Eddington
accretion onto an intermediate mass black hole despite many objects
having apparently cool disc temperatures. We argue that these observed
disc temperatures are not a good indicator of the black hole mass as
the powerful, optically thick corona drains energy from the inner
disc, and obscures it. We estimate the intrinsic (corona-less) disc
temperature, and demonstrate that in most cases it lies in the regime
of stellar mass black holes. These objects have spectra which range
from those similar to the highest mass accretion rate states in
Galactic binaries (a single peak at 2--3~keV), to those which clearly
have two peaks, one at energies below 1~keV (from the outer,
unComptonised disc) and one above 3~keV (from the Comptonised, inner
disc). However, a few ULXs have a significantly cooler corrected disc
temperature; we suggest that these are the most extreme stellar mass
black hole accretors, in which a massive wind completely envelopes the
inner disc regions, creating a cool photosphere.  We conclude that
ULXs provide us with an observational template for the transition
between Eddington and super-Eddington accretion flows, with the latter
occupying a new {\em ultraluminous} accretion state.

\end{abstract}


\section{Introduction} 

Ultraluminous X-ray sources (hereafter ULXs) are bright X-ray sources
with $L_{\rm X} > 10^{39}$~erg~s$^{-1}$. These objects are not
associated with the nuclei of galaxies, so are not powered by
accretion onto a central super-massive black hole, but are too bright
for sub-Eddington accretion onto stellar-mass black holes that radiate
isotropically.  An obvious solution is therefore an object
intermediate in mass between that of stellar mass and super-massive
black holes: intermediate mass black holes (IMBHs), of mass $\sim$
10$^2$ - 10$^4$~$M_{\odot}$, accreting at sub-Eddington rates
(e.g. Colbert \& Mushotzky 1999; Miller \& Colbert 2004).  Various
pathways for the formation of IMBHs have been suggested, both
primordial (from the collapse of Population III stars; Madau \& Rees
2001), and ongoing (runaway mergers of massive stars, and their
subsequent collapse to a massive black hole, in the densest
star-formation regions e.g. Portegies-Zwart et al. 2004), but problems
persist in producing sufficient numbers of IMBHs to tally with those
seen in the local Universe (Madhusudhan et al. 2006).  Thus an
alternative must be considered: that the majority of ULXs are single
stellar remnant black holes that accrete material from a close
companion star, probably at super-Eddington rates and/or with their
X-ray emission subject to some degree of geometric beaming (e.g. King et
al. 2001, Begelman 2002; see also Roberts 2007 and references
therein). More extreme beaming from relativistic jets seems unlikely 
due to the number of unbeamed sources that would need to be 
present in systems such as the Cartwheel galaxy (King 2004)

Either conclusion remains controversial, and the question of what
underlies ULXs can only be ultimately solved by a direct mass
measurement based on constraints placed by the binary orbit.  Until
this is achieved, more indirect methods of determining the mass must
be used in an attempt to constrain the nature of these accreting
systems. A common method is to use the temperature of the accretion
disc emission, together with its luminosity, to determine the mass as
Shakura-Sunyaev models predict the disc temperature $kT \sim
(M/10M_\odot)^{-1/4} (L/L_{\rm Edd})^{1/4}$ keV (Shakura \& Sunyaev
1973).  The highest signal-to-noise spectra of ULXs can often be well
fitted by composite models of a disc together with a hard ($\Gamma$
$<$ 2) tail, with the low measured disc temperature of $\sim$ 0.2~keV
implying a high black hole mass of $\sim$ 10$^3$ $M_\odot$
(e.g. Miller et al. 2003; Kaaret et al. 2003; Miller, Fabian \& Miller
2004). This supports the IMBH interpretation, and the resulting very
sub-Eddington accretion rates ($<$ 0.1$L_{\rm Edd}$) would most likely
be associated with the low/hard state in black hole binaries (BHBs),
consistent with the observed hard tail.

However, it is also clear from many studies of BHBs that disc models
only give reliable results when the X-ray spectrum is dominated by
this component (the high/soft or thermal dominant state) (Done \&
Kubota 2006). The derived disc temperature is increasingly distorted
as the tail increases in importance, with very low temperature, high
luminosity discs (implying a much larger mass black hole than is known
to be present!) seen when both tail and disc are strongest, as in the
very high or steep power law state (Kubota \& Done 2004, Done \&
Kubota 2006). The ULX spectral decompositions of Miller et al. (2004)
all have strong tails ($\ga 80\%$ of the 0.3 - 10 keV flux; Stobbart,
Roberts \& Wilms 2006, hereafter SRW06), so a straightforward
interpretation of the derived disc parameters is unlikely to give a
robust mass estimator\footnote{It should also be noted that band pass
is a potential problem in comparing ULX results with those of
BHBs. Given their high fluxes, Galactic sources have been most
commonly studied in detail using telescopes such as {\it RXTE\/}, so
are generally characterised by their behaviour in the 3--20 keV
regime.  The more distant (and hence fainter) ULXs instead require the
more sensitive, high spatial resolution telescopes of {\it Chandra\/}
and {\it XMM-Newton\/} for detailed studies, whose CCD detectors cover
the 0.3--10~keV band. This difference in typical energy range between
the ULX and BHB data mean that any comparison between the two must be
made with care.}.

However, the best data also shows the limitations of this simple
spectral fitting. There is clear evidence for curvature of the tail at
the highest energies, with a deficit of photons above 5~keV (Roberts
et al. 2005; SRW06; Miyawaki et al. 2009). Such curvature is never
seen in the low/hard state at these low energies. Only the very high
state shows curvature at such low energies, but these generally have
spectra which are steep.  The temporal variability also does not
correspond to that expected from an IMBH in this (or any other)
state. The simple disc and tail spectral decomposition shows that the
tail dominates the {\it XMM-Newton\/} band pass, so by analogy with
the BHBs this emission should be strongly variable. Yet many {\it
XMM-Newton\/} ULX light curves show only upper limits on the
variability power consistent with the poisson noise level for the data
(Feng \& Kaaret 2005).  Thus neither the spectral nor variability
properties convincingly correspond to any of the accretion states
known in BHBs, making it unlikely that ULXs are powered by
sub-Eddington flows onto an IMBH (Roberts 2007).


This conclusion is strongly supported by population studies of
ULXs. The sheer numbers observed in starburst galaxies together with
their short lifetimes (implied by their location in regions of active
star formation) mean that an unrealistically large underlying
population of IMBHs must be present (King 2004). Instead it is much
more likely that the bulk of the population are the most extreme
examples of High Mass X-ray Binaries (HMXBs) which somehow exceed the
Eddington limit. The high mass companion gives a natural origin for
the high mass transfer rates required to power the observed
luminosities (Rappaport, Podsiadlowski \& Pfahl 2005). It also
explains the association of ULXs with star-forming regions (Fabbiano,
Zezas \& Murray 2001; Lira et al. 2002; Gao et al. 2003) and the
unbroken luminosity function connecting ULXs to the standard X-ray
binary population (Grimm, Gilfanov \& Sunyaev 2003).  Indeed, direct
evidence for ULXs possessing high mass donor stars comes from the
identification of luminous, blue optical stellar counterparts to
several nearby ULXs (e.g. Liu et al. 2004; Kuntz et al. 2005; Roberts,
Levan \& Goad 2008).

Many ULX host galaxies have sub-solar abundances (e.g. Lee et al
2006). The lower opacity of subsolar material means that massive stars
lose less mass through winds during their evolution, so potentially
can collapse to form a relatively high mass black hole at the end of
their stellar lifetime, with M$_{\rm BH}$ $\la$ 80~$M_\odot$ (Fryer
\& Kalogera 2001; Heger et al. 2003; Belczynski, Sadowski \& Rasio
2004; Belczynski et al. 2009). One such large stellar mass black hole was discovered in IC 10
X-1, a Wolf-Rayet black hole binary. A radial velocity curve was
constructed from repeated optical observations, which provided a mass
estimate of 23-34 $M_\odot$ (Prestwich et al. 2007; Silverman \&
Filippenko 2008). Nonetheless, even such massive stellar remnant black
holes are required to be accreting at super-Eddington rates to explain
the observed luminosities of the brightest ULXs. Thus the accretion
flows in ULXs may not simply be scaled up versions of those seen in
BHBs, as would be the case for the IMBH\footnote{Here and throughout
the paper we distinguish IMBHs as BHs with masses $> 100 M_{\odot}$,
as suggested by the X-ray spectral modelling of Miller et al. (2003)
and subsequent work.  Such black holes are not formed at the endpoint
of a recent single stellar evolution, but require more exotic origins
such as stellar mergers in a young super star cluster
(e.g. Portegies-Zwart \& McMillan 2002) or formation from primordial
Population III stars in the high-redshift Universe (Madau \& Rees
2001).} model where the flows are sub-Eddington. Instead, observation
of ULXs may allow us to probe a new regime of accretion physics, a new
``ultraluminous state'' (Roberts 2007; Soria 2007).

Here we revisit the question of what the X-ray spectra of ULXs can
tell us about the nature of their accretion flows, and how this
constrains the nature of the accreting object, by utilising the best
data currently available in the {\it XMM-Newton} public archives.  We
choose to use the best available data as previous studies of samples
of ULXs (e.g. Berghea et al. 2008), whilst providing interesting
results, are ultimately limited in the conclusions they can draw by
the moderate signal-to-noise of many ULX spectra.  By using only the
highest quality data from the widest band pass, highest sensitivity
instruments available we can hope to avoid the ambiguity of previous
analyses, and make definitive statements on the accretion processes in
ULXs.  The paper is arranged as follows.  First we outline the sample
selection, and the data reduction processes
(Sections~\ref{section:selection} \& \ref{section:data}).  Next we
investigate the X-ray spectra using a variety of empirical and
physical models, in order to both clarify the morphology of ULX
spectra in the putative ultraluminous state, and to investigate the
physical processes underlying this phenomenon
(Section~\ref{section:spectral}).  Finally we discuss the implications
of our results for the nature of ULXs
(Section~\ref{section:discussion}).

\section{Source Selection}
\label{section:selection}

\renewcommand{\baselinestretch}{1.0} %
\begin{table*}
\leavevmode
\begin{center}
\caption{The ULX sample.}
\label{tab:sample}
\begin{tabular}{llccccc}
\hline
Source       & Alternative names       & RA (J2000)  & Dec. (J2000) & $N_{\rm H}$ $^a$             & $d$ $^b$           & $L_{\rm X}$ $^c$\\
                    &                                      &                       &                   & (10$^{20}$ cm$^{-2}$) & (Mpc)            & (10$^{39}$ erg s$^{-1}$) \\
\hline
NGC 55 ULX$^1$         & XMMU J001528.9-391319$^{2}$         & 00 15 28.9  & -39 13 19.1  & 1.71                  & 1.78({\it i})    & 1.1\\ 
                                          & NGC 55 6$^{3}$                                       &                       &                        &                           &                     &      \\
                                          & Source 7$^{4}$                                        &                       &                        &                           &                      &      \\
M33 X-8$^1$                  & CXOU J013351.0+303937$^5$          & 01 33 50.8  & +30 39 37.1  & 5.58                  & 0.70({\it i})    & 1.0\\
                                          & NGC 598 ULX1$^6$                              &                       &                        &                           &                      &      \\
                                          & Source 3$^7$                                          &                       &                        &                           &                      &      \\
NGC 1313 X-1$^1$      & IXO 7$^8$                                                 & 03 18 20.0  & -66 29 11.0  & 3.90                  & 3.70({\it i})     & 3.7\\
                                          & Source 4$^7$                                          &                       &                        &                           &                      &      \\
NGC 1313 X-2$^1$      & IXO 8$^8$                                                 & 03 18 22.3  & -66 36 03.8  & 3.90                  & 3.70({\it i})     & 4.7\\ 
                                          & NGC 1313 ULX3$^9$                            &                       &                        &                           &                      &      \\
                                          & Source 5$^7$                                          &                       &                        &                           &                      &      \\
IC 342 X-1$^{10}$        & CXOU J034555.7+680455$^{11}$      & 03 45 55.5  & +68 04 54.2  & 31.1                  & 3.3({\it ii})     & 2.8\\
                                          & IXO 22$^8$                                              &                       &                        &                           &                     &      \\
                                          & PGC 13826 ULX3$^9$                          &                       &                        &                           &                     &      \\
NGC 2403 X-1$^1$       & CXOU J073625.5+653540$^{12}$    & 07 36 25.6  & +65 35 40.0  & 4.17                  & 4.20({\it i})    & 2.4\\
                                          & Source 21$^{13}$                                   &                       &                        &                           &                     &      \\
                                          & NGC 2403 X2$^9$                                 &                       &                        &                          &                     &      \\
Ho II X-1$^1$                 & IXO 31$^8$                                               & 08 19 29.0  & +70 42 19.3  & 3.42                  & 4.50({\it i})  & 14.4 \\ 
                                          & PGC 23324 ULX1$^9$                           &                       &                        &                           &                     &      \\
                                          & CXOU J081928.99+704219.4$^{12}$ &                       &                        &                           &                      &      \\
M81 X-6$^1$                  & NGC 3031 ULX1$^6$                             & 09 55 32.9  & +69 00 33.3  & 4.16                  & 3.63({\it iii})   &  2.2\\ 
                                          & CXOU J095532.98+690033.4$^{12}$ &                       &                        &                           &                     &      \\
Ho IX X-1$^1$               & M81 X-9$^{1}$                                           & 09 57 53.2  & +69 03 48.3  & 4.06                  & 3.55({\it i}) & 7.5\\
                                          & NGC 3031 10$^{14}$                              &                       &                        &                           &                     &      \\
                                          & IXO 34$^8$                                                &                       &                        &                           &                     &      \\
                                          & H 44$^{15}$                                               &                       &                        &                           &                     &      \\
                                          & Source 17$^7$                                          &                       &                        &                           &                     &      \\
NGC 4559 X-1$^1$      & IXO 65$^8$                                                 & 12 35 51.7  & +27 56 04.1  & 1.49                  & 9.70({\it i}) & 8.0\\ 
                                          & CXOU J123551.71+275604.1$^{12}$  &                       &                        &                           &                     &      \\
                                          & X-7$^{16}$                                                  &                       &                        &                           &                     &      \\
NGC 5204 X-1$^1$      & IXO 77$^8$                                                  & 13 29 38.6  & +58 25 05.7  & 1.39                  & 4.80({\it i}) & 5.3\\
                                          & CXOU J132938.61+582505.6$^{12}$  &                       &                        &                           &                     &      \\
                                          & Source 23$^7$                                           &                       &                        &                           &                     &      \\
NGC 5408 X-1$^{17}$ & J140319.606-412259.572$^{17}$         & 14 03 19.6   & -41 22 59.6  & 5.67                  & 4.80({\it iv})    & 3.7 \\
                                          & Source 25$^7$                                           &                       &                        &                           &                     &      \\
\hline
\end{tabular}
\end{center}
\begin{minipage}{\textwidth}
Notes: $^a$Absorption column values taken from Dickey \& Lockman
(1990) using \textsc{webpimms}.  $^b$Figures shown in brackets relate
to following references, from which the assumed distance was taken:
({\it i}) SRW06, ({\it ii}) Saha et al. (2002), 
({\it iii}) Liu \& Di Stefano (2008), 
({\it iv}) Karachentsev et al. (2002).
$^c$observed X-ray luminosity (0.3--10.0 keV) based on the DKBBFTH model (see later).
Numbers shown in superscript relate to the following references for source names: $^1$SRW06, $^2$Stobbart et al. (2004), $^3$Read, Ponman \& Strickland (1997), 
$^4$Schlegel, Barrett \& Singh (1997), $^5$Grimm et al. (2005), $^6$Liu \& Mirabel (2005), 
$^7$Feng \& Kaaret (2005), $^8$Colbert \& Ptak (2002), $^9$Liu \& Bregman (2005), $^{10}$Roberts \& Warwick (2000),
$^{11}$Roberts et al. (2004), $^{12}$Swartz et al. (2004), $^{13}$Schlegel \& Pannuti (2003), $^{14}$Radecke (1997),
$^{15}$Immler \& Wang (2001), $^{16}$Vogler, Pietsch \& Bertoldi (1997), $^{17}$Kaaret et al. (2003). 
\end{minipage}
\end{table*}

Following the example of SRW06, we aim to use only the highest quality
data publicly available from the {\it XMM-Newton} Science archive
(XSA\footnote{See {\tt http://xmm.esac.esa.int/xsa/}}) in order to
provide the best characterisation of the structure of ULX spectra. We
therefore choose only the best data sets, ULX observations with $\ga
10,000$ accumulated EPIC counts ($\ga$ 500 independent spectral bins
available for fitting). This restriction is imposed based on the work
of SRW06, whose analysis shows that this is a reasonable threshold for
statistically distinguishing between physically motivated models,
particularly above 2 keV. This constraint provides a sample of 12
sources, which are listed in Table~\ref{tab:sample}. We note that this
may not be an exhaustive list, but that the number of sources in our
sample is probably sufficient to allow global trends to become
apparent.  Some of this sample of ULXs have been observed on more than
one occasion, in which case we select the longest individual exposure
to provide the clearest view of their spectrum. The selected ULXs all
reside within nearby galaxies ($\la$10 Mpc) due to restrictions
enforced by data quality, and vary in foreground Galactic absorption
in the range 1.39 -- 31.1 $\times$ 10$^{20}$ cm$^{-2}$.  Their X-ray
luminosities are representative of the full ULX range, $\sim$
10$^{39}$ -- a few 10$^{40}$ erg s$^{-1}$.

\section{Observations and Data Reduction}
\label{section:data}

Data from the longest individual observation of each source in our
sample were downloaded from the XSA. The data sets were reduced using
standard tools in {\it XMM-Newton|\/} \textsc{SAS} software (version
7.0.0)\footnote{See {\tt http://xmm.esac.esa.int/sas/}}. We found that
background flaring was severe enough in three cases that periods of
data were lost, resulting in multiple exposure data sets within the
same observation. Such flaring events took place during the
observations of Holmberg II X-1 and M81 X-6 and caused multiple
exposures in the MOS detectors only. In each case we find that one of
the MOS exposures was heavily contaminated, so for these objects we
only use the other MOS data, together with the pn data. The third
observation to be affected in this way is that of M33 X-8.  In this
case we find that the pn data is split into 2 exposures whilst data
from the MOS detectors is split into 3 exposures. The first exposure
from each detector contains no usable information.  On further
examination we find that the second MOS exposure is heavily
contaminated by flaring, we therefore only use the second pn and third
MOS exposures for the observation of this object. To remove any
remaining flaring events from our observations we constructed good
time interval (GTI) files from pn data using a full-field 10 -- 15 keV
background light curve and count rate criteria. The exact value of the
count rate criteria used to construct the source GTI files vary
according to field (typically excluding count rates higher than $\sim
1 - 1.5$ ct s$^{-1}$) to provide the longest exposure whilst
minimising contamination. Details of these observations are included
in Table~\ref{tab:obs}, with listed exposure times incorporating the
GTI corrections used during the reduction of the data.

\renewcommand{\baselinestretch}{1.0} %
\begin{table}
\leavevmode
\begin{minipage}{80 mm}
\begin{center}
\caption{Observation details.}
\label{tab:obs}
\begin{tabular}{lcccr}
\hline
Source       & Obs ID     & Date       & Off axis angle$^a$   & Exp$^b$ \\
                    &                   &                &   (arcmins)         & \multicolumn{1}{c}{(s)}          \\
\hline
NGC 55 ULX   & 0028740201 & 2001-11-14 & 4.2 & 30410 \\
M33 X-8            & 0102640101 & 2000-08-04 & 0.4 &  8650 \\
NGC 1313 X-1 & 0405090101 & 2006-10-15 & 0.8 & 90200 \\
NGC 1313 X-2 & 0405090101 & 2006-10-15 & 6.2 & 90200 \\
IC 342 X-1        & 0206890201 & 2004-08-17 & 3.6 & 19750 \\
NGC 2403 X-1 & 0164560901 & 2004-09-12 & 5.4 & 58470 \\
Ho II X-1            & 0200470101 & 2004-04-15 & 0.3 & 40800 \\
M81 X-6            & 0111800101 & 2001-04-22 & 3.2 & 88300 \\
Ho IX X-1          & 0200980101 & 2004-09-26 & 0.3 & 80400 \\
NGC 4559 X-1 & 0152170501 & 2003-05-27 & 0.3 & 38300 \\
NGC 5204 X-1 & 0405690201 & 2006-11-19 & 0.2 & 33700 \\
NGC 5408 X-1 & 0302900101 & 2006-01-13 & 0.2 & 99300 \\
\hline
\end{tabular}
\end{center}
Notes: $^a$Off axis angle of source in the {\it XMM-Newton} EPIC field of view, $^b$Sum of good
time intervals for each observation (taken from pn data), calculated
as per in the text.
\end{minipage}
\end{table}

The source spectra were extracted from circular apertures centred on
the individual ULXs in each detector. This was straightforward for the
majority of data, but we found that NGC 1313 X-2 was unfortunately
positioned on the chip gap of the MOS1 detector so a polygonal source
region was applied to optimise our data extraction. Background spectra
were obtained from larger circular regions placed near to the
source. Where possible, these were positioned on the same chip and at
a similar distance from the read out node as the source. The only
exception was NGC 4559 X-1, where the MOS detectors were operating in
small window mode, therefore alternative background regions were
selected on a separate chip in a position as close as possible to that
used in the pn detector, whilst in the case of M81 X-6 no data was
contained in the MOS1 observation. The size of the individual source
and background extraction regions are recorded in Table
\ref{tab:extract} (in the case of NGC 1313 X-2 we only list the size
of the circular apertures used in pn and MOS2).

\renewcommand{\baselinestretch}{1.0} %
\begin{table}
\leavevmode
\begin{minipage}{85 mm}
\begin{center}
\caption{Size of regions used to extract the spectral data, and details of the resultant spectra.}
\label{tab:extract}
\begin{tabular}{lcccc}
\hline
Source       & \multicolumn{2}{c}{Extraction radius}                        & Spectral bins$^a$ & Rate$^b$   \\
	& \multicolumn{2}{c}{(arcseconds)}	& & (ct s$^{-1}$) \\
             &  \multicolumn{1}{c}{Source} & \multicolumn{1}{c}{Background} &               & \\
\hline
NGC 55 ULX   & 34 & 51   & 884  & 2.13 \\
M33 X-8            & 50 & 75   & 1252 & 9.56 \\
NGC 1313 X-1 & 40 & 60   & 1616 & 1.19 \\
NGC 1313 X-2 & 40$^*$ & 60   & 1600 & 1.04 \\
IC 342 X-1        & 36 & 54   & 516  & 0.68 \\
NGC 2403 X-1 & 35 & 52.5 & 843  & 0.48 \\
Ho II X-1            & 52 & 78   & 1358 & 4.93 \\ 
M81 X-6            & 22 & 33   & 989  & 0.69 \\
Ho IX X-1          & 42 & 63   & 2139 & 2.49 \\
NGC 4559 X-1 & 34 & 51   & 593  & 0.50 \\ 
NGC 5204 X-1 & 40 & 60   & 873  & 1.55 \\
NGC 5408 X-1 & 40 & 60   & 990  & 1.42 \\
\hline
\end{tabular}
\end{center}
Notes: $^a$Number of spectral bins available for fitting from combined
EPIC detectors; $^b$combined EPIC count rate of source during
observation in the 0.3--10 kev band. $^*$Alternative extraction region
used in MOS1 because source positioned on edge of chip gap.
\end{minipage}
\end{table}

The best quality data (\textsc{FLAG} = 0) were extracted in each case
with \textsc{PATTERN} $\leq$ 4 for pn and \textsc{PATTERN} $\leq$ 12
for MOS. The response and ancillary response files were created
automatically by the standard {\it XMM-Newton} tasks and spectral
files were grouped to a minimum of 20 counts per bin, to improve
statistics. The number of independent spectral bins after this process
was completed are listed in Table \ref{tab:extract}, along with the
combined EPIC source count rates.

\section{ULX Spectral Properties} %
\label{section:spectral}

Our aims in this work are twofold.  Firstly we aim to investigate the
basic shape of the ULX X-ray spectra, and in doing so evaluate the
evidence for the presence of a soft excess and a power-law break (at
energies of a few keV) in ULX spectra.  As these have been suggested
as the two distinguishing spectral features of a new, ultraluminous
accretion state (Roberts 2007) it is important to examine the evidence
for their presence in the highest quality ULX spectra offered by {\it
XMM-Newton\/}, and so determine the validity of claims of a new
accretion state.  Secondly, we will investigate the physical insights
that a range of models can afford us on the nature of the accretion
flows in this putative state.  The models we use vary from the
simplest empirical models (power-law continua), through accretion disc
models, to models in which we consider both an accretion disc and a
Comptonising corona surrounding its inner regions, and the interplay
between the two.


All spectra are fit in \textsc{xspec} version 11.3.2 over the
0.3--10.0~keV energy range (unless otherwise stated). To maximize data
quality, pn and MOS data were fit simultaneously, with the addition of
a constant multiplicative factor to compensate for calibration
difference between the cameras. The pn constant is fixed at unity,
whilst those for each MOS camera remain free, with the fitted values
generally agreeing to within $\sim$10~\% (larger discrepancies only
occurred with disparate extraction regions; see Section
\ref{section:data} above for more details).  In each case, the spectra
are fit with two absorption components; one fixed at the column
observed along the line of sight within our own Galaxy as listed in Table
\ref{tab:sample} (from Dickey \& Lockman 1990), and a second component
that is allowed to vary to represent any absorption within the host
galaxy and/or intrinsic to the ULX. The absorption columns are modeled
using the TBABS model (Wilms, Allen \& McCray 2000).  All quoted
errors are the 90\% confidence interval for one interesting parameter.

\subsection{Single component phenomenological models}
\label{subsection:single}

Simple, single component models have been used rather successfully to
describe the featureless spectra of ULXs in the low-to-moderate
quality data regime for some years (e.g. Humphrey et al. 2003; Swartz
et al. 2004; Feng \& Kaaret 2005; Winter, Mushotzky \& Reynolds 2006).
Here we use only the highest quality data to perform our analysis, yet
we still find the the X-ray spectra of these objects to appear
relatively smooth and featureless. Therefore, the application of these
simple, single continuum models is a good starting point in their
analysis. We start by applying an absorbed power-law continuum (PO in
\textsc{xspec} syntax) and an absorbed multi-coloured disc blackbody (hereafter
MCD) component (DISKBB in \textsc{xspec} ; Mitsuda et al. 1984)
separately to the data. The resultant fits can be seen in
Table~\ref{tab:single}.  

\begin{figure}
\begin{minipage}{85mm}
\begin{center}
\leavevmode
\epsfxsize=8cm \rotatebox{0}{\epsfbox{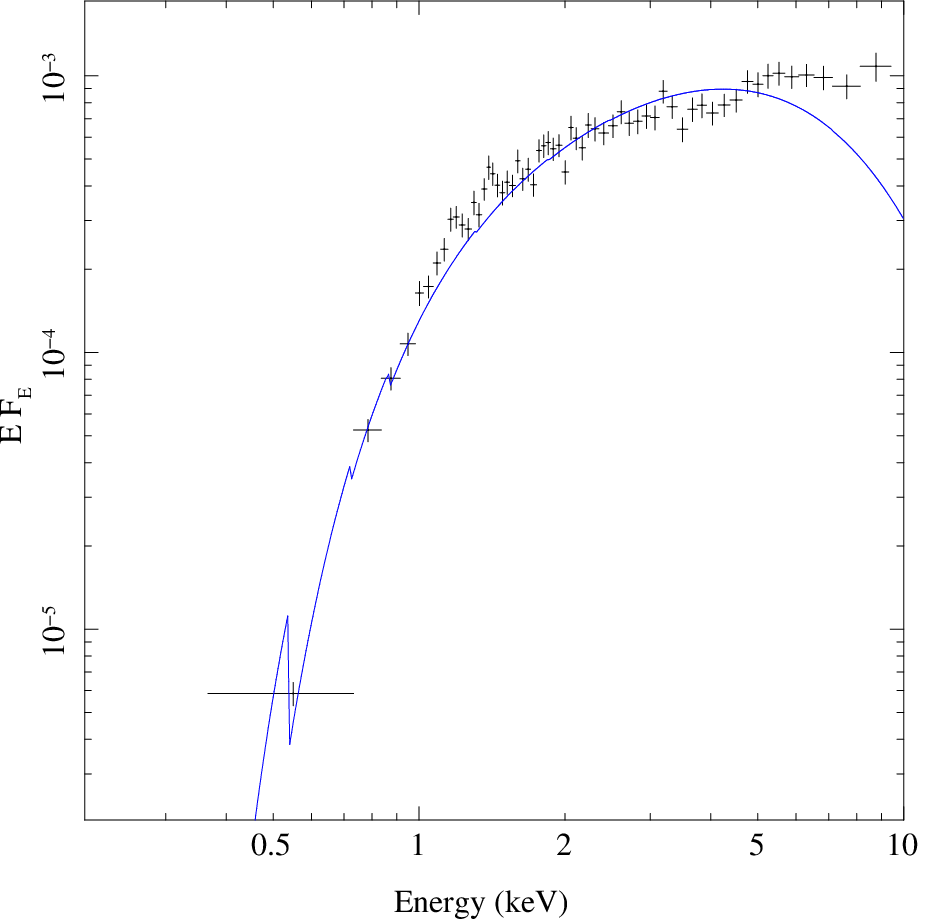}} 
\end{center}
\caption{{\it XMM-Newton} EPIC pn data from IC 342 X-1, fit with an
absorbed multi-colour disc model (DISKBB).  We plot only the
pn data, shown in black, which we rebin to a minimum of $10\sigma$ statistical significance, or 10
channels per data point for clarity.  Although this is statistically one of the
better fits to a MCD model within our sample, a visual inspection
quickly reveals the residuals (particularly evident at high energies)
that explain why it is still rejected at high significance.}
\label{fig:diskbb}
\end{minipage}
\end{figure}

\renewcommand{\baselinestretch}{1.5} 
\begin{table} 
\leavevmode 
\begin{minipage}{85 mm}
\begin{center} 
\caption{Simple power-law continuum and MCD spectral fits.} 
\label{tab:single} 
\begin{tabular}{l c c c}
\hline 
 Source & \multicolumn{2}{c}{Model parameters} & $\chi$$^2$/DoF$^{\dagger}$ \\
\hline
{\it TBABS*TBABS*PO}              & \multicolumn{1}{c}{$N_{\rm H}$$^a$}      & \multicolumn{1}{c}{$\Gamma$$^b$} & \\ 
\\
NGC 55 ULX    &  0.455$\pm$0.008   & 3.30$\pm$0.03    & 1288.8/879  \\ 
M33 X-8            &  0.249$^*$                 & 2.218$^*$            & 2527.3/1247    \\ 
NGC 1313 X-1 &  0.188$\pm$0.005   & 1.85$\pm$0.02    & 2138.5/1611     \\ 
NGC 1313 X-2 &  0.341$\pm$0.008   & 1.81$\pm$0.02    & 1982.4/1595     \\ 
IC 342 X-1        &  0.57$\pm$0.04        & 1.83$\pm$0.05    & 534.3/511    \\ 
NGC 2403 X-1 &  0.51$\pm$0.02       & 2.38$\pm$0.03     & 1247.3/838     \\ 
Ho II X-1            &  0.158$\pm$0.003  & 2.63$\pm$0.01     & 1602.6/1363\\ 
M81 X-6            &  0.39$\pm$0.01       & 2.09$\pm$0.02     & 1825.9/985   \\ 
Ho IX X-1          &  0.105$\pm$0.003  & 1.606$^{+ 0.01}_{- 0.009}$ & 2863.4/2112  \\ 
NGC 4559 X-1 & 0.120$\pm$0.009   & 2.29$\pm$0.04     & 586.5/588   \\ 
NGC 5204 X-1 &  0.153 $\pm$0.006 & 2.52$\pm$0.03     & 986.3/868    \\ 
NGC 5408 X-1 &  0.091$^{+ 0.004}_{- 0.003}$ & 3.12$\pm$0.02 & 1801.9/985    \\ 
\\
\\
{\it TBABS*TBABS} & \multicolumn{1}{c}{$N_{\rm H}$$^a$} & \multicolumn{1}{c}{$kT_{\rm in}$$^c$} & \\
 \multicolumn{1}{c}{{\it *DISKBB}} &                     &                                                 &                      \\
NGC 55 ULX   &  0.123$^*$                                  & 0.573$^*$                             & 2067.2/879   \\ 
M33 X-8            & 0.007$\pm$0.003                     & 1.11$\pm$0.01                    & 1470.7/1247   \\ 
NGC 1313 X-1 &  0.008$^*$                                 & 1.420$^*$                             & 5531.5/1611   \\ 
NGC 1313 X-2 & 0.115$\pm$0.004                     & 1.53$\pm$0.02                    & 2091.5/1595   \\ 
IC 342 X-1        &  0.19$\pm$0.02                         & 1.75$^{+ 0.07}_{- 0.06}$ & 775.1/511  \\ 
NGC 2403 X-1 &  0.168$^{+ 0.01}_{- 0.009}$   & 1.06$\pm$0.02                    & 891.0/838     \\ 
Ho II X-1            &  0.0$^*$                                      & 0.580$^*$                            & 8242.0/1363   \\ 
M81 X-6            & 0.111$\pm$0.006                     & 1.31$\pm$0.02                    & 1210.1/985    \\ 
Ho IX X-1          &  0.0$^*$                                      & 1.624$^*$                             & 9164.3/2112  \\ 
NGC 4559 X-1 & 0.0$^*$                                       & 0.708$^*$                             & 1545.6/588   \\ 
NGC 5204 X-1 &  0.0 $^*$                                     & 0.618$^*$                             & 2533.2/868   \\ 
NGC 5408 X-1 & 0.0$^*$                                       & 0.314$^*$                             & 8294.8/985    \\ 
\hline 
\end{tabular} 
\end{center} 
Notes: Model is abbreviated to \textsc{xspec} syntax: TBABS -
absorption components for both Galactic and external absorption; PO -
power-law; DISKBB - MCD. Specific notes: $^a$External
absorption column ($\times$ 10$^{22}$ cm$^{-2}$) left free during
fitting, Galactic columns listed in Table \ref{tab:sample}; $^b$
power-law photon index;  $^c$ inner-disc temperature (keV).  $^*$Best fitting models to this data give a
reduced $\chi$$^2$ greater than 2, hence we do not place constraints due to the paucity of the fit.  $^{\dagger}$Here and elsewhere we
use 'DoF' to abbreviate the number of degrees of freedom available
when fitting a model.
\end{minipage} 
\end{table}

Table \ref{tab:single} shows that a single power-law is not a
particularly good fit to ULX data of this high quality.  Although a
rough representation of the observed spectra (reduced $\chi^2$,
$\chi^2_{\nu} < 2$) is found for 11/12 objects, only two objects
provide statistically acceptable fits (null hypothesis probability $>
5\%$).  Notably, these two objects - IC 342 X-1 and NGC 4559 X-1 -
also have the worst quality data in the sample.  The vast majority of
the other ULX data reject this model at very high significance.  The
situation is even worse when fitting the MCD model, where the quality
of the fits are so bad they can only roughly represent five of the
twelve spectra, and only one data set (NGC 2403 X-1) does not reject
this model at high significance.

Nonetheless, it is instructive to compare the parameters derived from
these fits, since lower quality data would not be able to show the
inadequacy of these models. For the spectra that are best represented
by an absorbed MCD (again using the $\chi^2_{\nu} < 2$ criterion) we
find that $kT_{\rm in}$ $\sim$ 1.06 -- 1.7 keV. These are close to
those seen in high mass accretion rate Galactic sources in the
high/soft (thermal dominated) state (e.g. McClintock \& Remillard
2006), which if taken at face value would indicate that the ULXs
contain standard stellar mass black holes accreting at fairly high
mass accretion rates. Conversely, with a power-law model,
Table~\ref{tab:single} shows that the photon index measured when these
objects are represented by an absorbed power-law range from 1.6 $<$
$\Gamma$ $<$ 3.3. While the steepest spectra would correspond to the
very high (steep power-law) state, those with with $\Gamma$ $<$ 2.1
correspond instead to the low/hard state (McClintock \& Remillard
2006). The low/hard state is only seen at very sub-Eddington mass
accretion rates ($< 0.1 L/L_{\rm Edd}$), so a na{\"i}ve interpretation
of this spectral decomposition suggests that we are observing at least
some IMBHs.

Clearly the fits from these models contradict one another as
individual ULXs cannot be both stellar and IMBH -- we note that the
ULX data that is best represented by MCD fits generally also fits with
power-law continua with $\Gamma < 2.3$. It is therefore evident that
problems can arise when we attempt to draw physical meaning from
simple phenomenological models applied to ULX data.  In the case of
the observations in our sample we have sufficient data quality to
demonstrate that the underlying spectrum is more complex than either
standard disc or power-law models.  As we have no reason to think our
ULX sample is atypical of the class, then by extension this should
apply to all ULXs, if sufficient data were available for them.  This
should render any physical conclusions drawn from simple model fits to
low-to-moderate quality ULX data as suspicious.

Single component phenomenological models do not provide enough
flexibility in fitting to accurately constrain the majority of the
data, nor do they consistently provide a physically realistic basis on
which to interpret the data.  As our next step, we therefore turn to
more complex phenomenological models that have been used to
characterise the spectra of BHBs and ULXs.

\subsection{Combined phenomenological models}
\label{subsection:double}

\begin{figure}
\begin{minipage}{85mm}
\begin{center}
\leavevmode
\epsfxsize=7.85cm \rotatebox{0}{\epsfbox{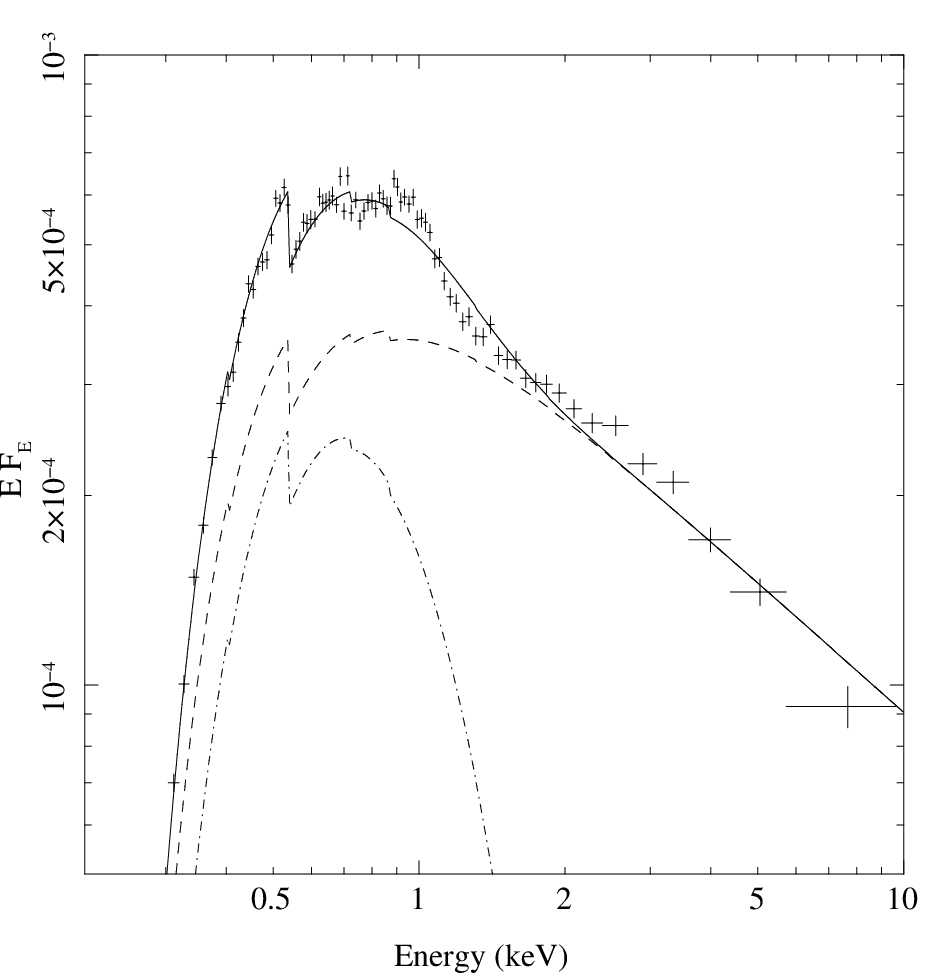}} 
\end{center}
\caption{{\it XMM-Newton\/} EPIC pn data from NGC 5408 X-1 (black data
points, rebinned to a minimum of 30$\sigma$ significance or 30 channels), shown with the best fitting
multi-colour disc plus power-law model (solid line).  The contributions
of the separate model components to the overall fit are overplotted as grey dashed
(power-law continuum) and dashed \& dotted lines (MCD).  This clearly shows the motivation for
an additional soft component in the spectrum.}
\label{fig:5408}
\end{minipage}
\end{figure}

\renewcommand{\baselinestretch}{1.0} %
\begin{table*}
\leavevmode
\begin{center}
\caption{Combined power-law plus MCD spectral fits.}
\label{tab:soft excess}
\begin{tabular}{lccccc}
\hline
Source      & \multicolumn{4}{c}{TBABS*TBABS*(PO+DISKBB)}   & $\Delta$$\chi$$^2$ $^d$ \\ 
 & \multicolumn{1}{c}{$N_{\rm H}$$^a$} & \multicolumn{1}{c}{$\Gamma$$^b$} & \multicolumn{1}{c}{$kT_{\rm in}$$^c$} & \multicolumn{1}{c}{$\chi$$^2$/DoF} & \\
\hline
NGC 55 ULX    & 0.46$\pm$0.02                        & 3.7$\pm$0.1                    & 0.77$^{+0.03}_{-0.04}$        & 1042.7/877    & 246.06 \\ 
M33 X-8            & 0.09$\pm$0.02                        & 2.0$\pm$0.1                     & 1.05$^{+0.03}_{-0.04}$       & 1216.0/1245  & 1311.30 \\ 
NGC 1313 X-1 & 0.26$^{+0.02}_{-0.01}$         & 1.70$^{+0.03}_{-0.02}$ & 0.23$\pm$0.01                       & 1796.5/1609 & 341.88 \\ 
NGC 1313 X-2 & 0.29$^{+0.03}_{-0.02}$         & 2.0$^{+0.2}_{-0.1}$         & 1.7$\pm$0.1                           & 1599.7/1593 & 382.77 \\ 
IC 342 X-1        & 0.7$^{+0.2}_{-0.1}$                & 1.7$\pm$0.1                      & 0.32$^{+0.1}_{-0.09}$          & 518.0/509     & 16.29 \\ 
NGC 2403 X-1 & 0.38$^{+0.08}_{-0.07}$         & 2.9$^{+0.3}_{-0.4}$         & 1.12$^{+0.04}_{-0.05}$        & 853.7/836     & 393.69 \\ 
Ho II X-1            & 0.116$\pm$0.007                   & 2.42$\pm$0.04                 & 0.37$\pm$0.02                      & 1503.9/1361 & 98.71 \\ 
M81 X6              & 0.30$^{+0.05}_{-0.04}$         & 2.6$^{+0.4}_{-0.3}$         & 1.42$^{+0.04}_{-0.05}$        & 1093.9/983   & 732.01 \\ 
Ho IX X-1           & 0.135$\pm$0.007                  & 1.46$\pm$0.02                 & 0.27$^{+0.02}_{-0.01}$        & 2440.0/2110 & 423.44 \\ 
NGC 4559 X-1 &  0.16$^{+0.03}_{-0.02}$        & 2.14$^{+0.07}_{-0.05}$  & 0.17$\pm$0.02                       & 528.1/586     & 58.36 \\ 
NGC 5204 X-1 &  0.09$\pm$0.01                       & 2.18$^{+0.08}_{-0.09}$  & 0.39$\pm$0.02                       & 925.1/866     & 61.16 \\ 
NGC 5408 X-1 &  0.068$^{+0.005}_{-0.006}$ & 2.68$\pm$0.04                 & 0.186$^{+0.007}_{-0.003}$ & 1320.5/983   & 481.41 \\ 
\hline 
\end{tabular}
\end{center}
\begin{minipage}{\textwidth}
Notes: models are abbreviated \textsc{xspec} syntax, as per Table~\ref{tab:single}. Specific notes:
$^a$External absorption column in units of 10$^{22}$ atoms cm$^{-2}$;
$^b$power-law photon index; $^c$inner disc temperature (keV); $^d$$\chi$$^2$
improvement over the absorbed power-law fit (see
Table~\ref{tab:single}), for two extra degrees of freedom. 
\end{minipage}
\end{table*}

Here we combine the two simplest continuum models to further
characterise the spectral shape and features of these sources. This
combination of a disc component plus a power-law has been used
extremely effectively in BHBs (McClintock \& Remillard 2006 and
references therein), since although these models are not terribly
physical, they provide a good approximation to a disc with an
optically thin Comptonising corona over the 3--20 keV range.  It has
also, of course, been used as the basis for claims of IMBHs underlying
ULXs (see introduction).  Each source spectrum is initially fit by an
absorbed power-law component, then a multi-coloured disc component is
added to this to look for any improvement of fit.  The results of
these fits can be seen in Table~\ref{tab:soft excess}.

Figure \ref{fig:5408} illustrates the need for a second spectral
component in ULX spectra. Here we see the spectrum of NGC 5408 X-1
deconvolved with the best fitting absorbed MCD plus power-law model.
It is clear from this data that a soft component is present in the
spectrum, as an excess above a harder continuum.  In fact, we find
that all our sources show a significant improvement in $\chi^2$ with
the addition of a disc component (Table \ref{tab:soft excess}).  This
implies that some form of soft excess is ubiquitous in ULX spectra at
this level of data quality.

Interestingly, in only 7/12 cases is the resulting disc `cool'
i.e. with $kT_{\rm in}<0.4$~keV as in Miller et al. (2004). Instead we
find that the temperature from the disc component appears to vary over
a much wider range, from 0.17 $<$ $kT_{\rm in}$ $<$ 1.7 keV, covering
both the cool and standard disc temperature range.  However, where the
disc is hot -- and so forms the predominant component at higher
energies -- we find that the power-law component contributes the soft
excess, agreeing with previous work (e.g. Stobbart, Roberts \& Warwick
2004; Foschini et al. 2004; Roberts et al. 2005).  As with these
analyses, we note that the power-law extends to lower energies than
its putative seed photons. This is not physically realistic and so
requires more plausible physical modelling.

Inspection of the residuals from these fits also reveals a possible
break or turn over above $\sim$ 2 keV. In some cases this is also
clearly visible in their spectra. We can see in Figure \ref{fig:5408}
that the power-law cannot model such a feature and so is plotted in an
average position whilst the data curves around it. Such a feature is
not present in standard BHB states and so deserves further
investigation, which is carried out in the next section.

Another point worthy of note is that the object with the highest
inner-disc temperature is NGC 1313 X-2, a ULX that has exhibited much
cooler disc temperatures in previous observations (e.g. 0.16~keV,
Miller et al. 2003) leading to the suggestion it harbours an IMBH. We
note that a local minimum in $\chi^2$ was observed at $\sim$~0.16~keV
in our data, but that the global minimum in our best fit occurs at
much higher temperatures ($kT_{\rm in}$ $=$ 1.7 keV). It is clear that
these results are contradictory -- NGC 1313 X-2 cannot change from an
IMBH to a stellar mass BHB between observations -- demonstrating yet
again that drawing physical conclusions from this soft component can
be hazardous and should therefore be approached with care.

\subsection{A high energy break}
\label{subsubsection:break}

\begin{figure}
\begin{minipage}{85mm}
\begin{center}
\leavevmode
\epsfxsize=8cm \rotatebox{0}{\epsfbox{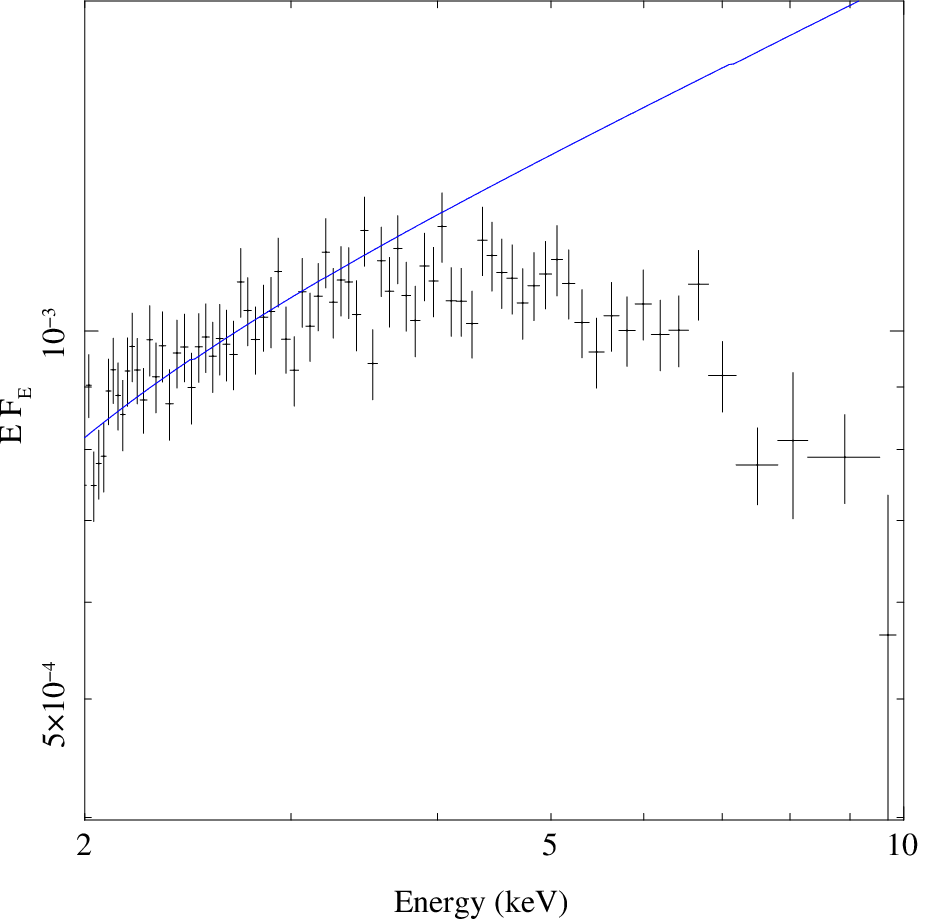}} 
\end{center}
\caption{{\it XMM-Newton\/} EPIC pn data from NGC 1313 X-2, displayed as per previous figures (rebinned to a minimum of 15$\sigma$, or 15 channels).  We show the data fit by a power-law component based on the slope and normalisation of the low energy (pre-break) component of the broken power-law fit to this data. It is very clear from this data that some form of break or curvature is present above 2 keV.}
\label{fig:55}
\end{minipage}
\end{figure}

\renewcommand{\baselinestretch}{1.0} %
\begin{table*}
\leavevmode
\begin{center}
\caption{A comparison of power-law to broken power-law spectal fits in the 2--10 keV band.}
\label{tab:break}
\begin{tabular}{lcccccccc}
\hline
Source        &  \multicolumn{2}{c}{PO}                 & \multicolumn{4}{c}{BKNPOWER}                                                        & $\Delta$$\chi$$^2$ $^e$ & 1-P({\it F}-test)$^f$ \\
                             & $\Gamma$$^a$      & $\chi$$^2$/DoF    & $\Gamma_1$$^b$        & $E_{\rm break}$$^c$     & $\Gamma_2$$^d$        & $\chi$$^2$/DoF &  \\ 
\hline
NGC 55 ULX     & 3.57$\pm$0.06               & 429.8/314       & 3.1$\pm$0.1                    & 3.9$\pm$0.3             & 4.9$^{+0.5}_{-0.4}$ & 323.1/311      & 106.7 & $>$99  \\
M33 X-8             & 2.60$\pm$0.03                & 855.6/679      & 2.17$^{+0.08}_{-0.1}$   & 4.0$^{+0.2}_{-0.3}$ & 3.4$\pm$0.2             & 641.6/677      & 214.0 & $>$99  \\
NGC 1313 X-1  & 1.06$\pm$0.02                & 1054.2/1043 & 1.60$^{+0.03}_{-0.05}$ & 6.3$^{+0.3}_{-0.9}$ & 2.6$^{+0.3}_{-0.5}$ & 992.9/1041   & 61.3   & $>$99  \\
NGC 1313 X-2  & 1.91$\pm$0.02                & 1195.8/1034 & 1.53$^{+0.1}_{-0.07}$   & 3.7$^{+0.7}_{-0.2}$ & 2.3$^{+0.2}_{-0.1}$ & 1011.0/1032 & 184.8 & $>$99  \\
IC 342 X-1         & 1.58$^{+0.06}_{-0.05}$ & 281.3/269      & 1.53$\pm$0.07                & 6.7$^{+0.7}_{-1.0}$ & 2.7$^{+1}_{-0.8}$    & 273.7/267      & 7.6      & $>$99  \\
NGC 2403 X-1  & 2.67$^{+0.05}_{-0.06}$ & 478.1/335      & 2.1$\pm$0.1                    & 4.0$\pm$0.2              & 4.0$\pm$0.3             & 335.0/333      & 147.1 & $>$99  \\
Ho II X-1             & 2.58$^{+0.02}_{-0.03}$ & 772.7/795      & 2.51$\pm$0.04               & 5.4$^{+0.5}_{-0.6}$  & 3.1$^{+0.3}_{-0.2}$ & 750.5/793      & 22.2   & $>$99  \\
M81 X-6             & 2.31$\pm$0.03                & 903.9/538      & 1.72$^{+0.07}_{-0.1}$   & 4.1$\pm$0.2              & 3.4$\pm$0.2             & 554.9/536      & 349.0 & 100  \\
Ho IX X-1           & 1.46$\pm$0.02                & 1672.6/1544 & 1.38$\pm$0.02               & 6.2$^{+0.3}_{-0.4}$  & 2.2$\pm$0.2             & 1561.1/1542  & 111.5 & $>$99  \\
NGC 4559 X-1  & 2.22$^{+0.07}_{-0.09}$ & 141.1/153      & 2.1$\pm$0.1                   & 4.8$^{+1}_{-0.9}$      & 2.8$^{+0.8}_{-0.4}$ & 133.8/151      & 7.3      & 98.1 \\
NGC 5204 X-1  & 2.40$\pm$0.06                & 319.5/307      & 2.36$^{+0.09}_{-0.2}$  & 5$^{+3}_{-2}$            & 2.7$^{+2}_{-0.5}$     & 316.7/305      & 2.8      & 73.9 \\
NGC 5408 X-1  & 2.84$^{+0.04}_{-0.06}$ & 442.0/417      & 2.80$\pm$0.06               & 7.1$^{+1}_{-0.9}$     & 7$^{+7}_{-2}$            & 432.4/415      & 9.6     & 98.9 \\
\hline
\end{tabular}
\end{center}
\begin{minipage}{\textwidth}
Notes: models are abbreviated to \textsc{xspec} syntax: PO - as before; BKPOWER - broken power-law model.  Specific notes:  $^a$Photon index from PO model, $^b$photon index before the break in BKNPOWER model, $^c$break energy (keV), $^d$photon index after the break energy,  $^e$$\chi$$^2$ improvement over a single power-law fit, $^f$ statistical probability (in per cent) that the broken power-law model provides an improvement to the fit over a single PO model, from the F-test.
\end{minipage}
\end{table*}

We investigate the prevalance of a high energy downturn following the
analysis of SRW06, i.e.  by comparing a power-law and a broken
power-law description of the data above 2~keV. We do not include
absorption in these fits as it is not easily constrained by data above
2 keV, and besides the wider band fits infer columns that have little
effect above 2 keV. The resulting fits are shown in Table
\ref{tab:break}.


The broken power-law is statistically preferred ($>$ 98 \%
significance improvement in fit according to the F-test) in eleven out
of the twelve ULXs in our sample, with break energies in the $\sim 3.5
- 7$~keV range, and a typical steepening of the power-law slope by
$\Delta \Gamma \sim 1 - 2$.  This near ubiquity is made all the more
remarkable when it is considered that the one ULX without evidence for
a break here -- NGC 5204 X-1 -- has displayed evidence for such a
feature in previous observations (Roberts et al. 2005, SRW06).  While
this break is clearly expected from the `hot disc' fits, where the MCD
component dominates at high energies, the break is also seen at high
significance in the `cool disc' objects. These are the sources where
the IMBH model is apparently favoured, implying that these ULX are
analogous to the low/hard state observed in Galactic BHBs. Yet the
high energy data show a break at $\sim$~5~keV which is not seen this
accretion state. The ubiquity of the high energy break shows that the
ULX spectra are {\em not} well described by a cool disc plus power-law
as expected from the IMBH model.

Instead, these high quality data show that the disc is either hot,
implying a stellar mass black hole (which must be accreting at or
above Eddington in order to produce the observed luminosity) but with
a previously unseen soft excess (an alternative would be that the power-law is simple broadening the disc spectrum to explain a known state; see later for more detailed discussions), or the disc is cool but with a high
energy tail which is quite unlike that seen in the standard spectral
states of BHBs (e.g. McClintock \& Remillard 2006). {\bf This new
combination of observational characteristics - a cool disc and a
broken harder component - suggests that these ULXs are operating in an
accretion state not commonly seen in the Galactic BHBs}. We term this
new combination of observational characteristics the {\bf ``ultraluminous
state''}, and it seems most straightforward to assume that this state
accompanies extremely high accretion rates onto stellar remnant black
holes.  As we also have no reason to think our ULX sample atypical of
the class as a whole, we can only presume that this spectrum may be
endemic to the majority of ULXs.

\renewcommand{\baselinestretch}{1.5}
\begin{table*}
\leavevmode
\begin{center}
\caption{Spectral fits for the $p$-free disc model.}
\label{tab:diskpbb}
\begin{tabular}{l c c c c}
\hline
Source               & \multicolumn{4}{c}{TBABS*TBABS*DISKPBB}    \\
                            & \multicolumn{1}{c}{N$_H$$^a$} & \multicolumn{1}{c}{T$_{in}$$^b$} &  \multicolumn{1}{c}{$p$$^c$}    &  $\chi$$^2$/DoF               \\ 
\hline
NGC 55 ULX    &  0.38 $\pm$ 0.01                    & 1.12$^{+0.06}_{-0.05}$  & 0.408$\pm$0.005                   & 1019.9/878     \\
M33 X-8            &  0.090$^{+0.009 }_{-0.01}$  & 1.36$^{+0.03}_{-0.04}$  & 0.599$^{+0.01}_{-0.009}$     & 1242.0/1246  \\
NGC 1313 X-1 &  0.188$^{+0.003}_{-0.005}$ & 7.9$\pm$0.8                     & 0.521$\pm$0.002                   & 2113.2/1610   \\
NGC 1313 X-2 &  0.251$^{+0.009}_{-0.01}$   & 2.31$^{+0.09}_{-0.1}$    & 0.583$^{+0.009}_{-0.006}$  & 1601.2/1594  \\
IC 342 X-1        &  0.57$\pm$0.04                       & 13$^{+7}_{-4}$                & 0.523$\pm$0.007                   & 532.1/510       \\
NGC 2403 X-1 &  0.27$\pm$0.03                       & 1.24$\pm$0.06                & 0.60$^{+0.03}_{-0.02}$         & 850.5/837       \\
Ho II X-1            &  0.149$\pm$0.003                   & 2.80$^{+ 0.1}_{-0.10}$  & 0.436$\pm$0.001                   & 1557.0/1362  \\
M81 X-6            &  0.23$^{+0.01}_{-0.02}$         & 1.60$^{+0.05}_{-0.06}$ & 0.61$^{+0.02}_{-0.01}$         & 1091.7/984   \\
Ho IX X-1          &  0.103$\pm$0.003                   & 8.9$^{+2}_{-0.4}$           & 0.557$^{+0.001}_{-0.002}$  & 2853.2/2111 \\
NGC 4559 X-1 &  0.121$^{+0.01}_{-0.009}$    & 6.2$^{+0.6}_{-0.8}$        & 0.466$^{+0.004}_{-0.002}$  & 574.7/587      \\
NGC 5204 X-1 &  0.147$^{+0.007}_{-0.004}$ & 3.0$\pm$0.2                     & 0.446$^{+0.002}_{-0.003}$  & 977.5/867        \\
NGC 5408 X-1 &  0.085$\pm$0.004                  & 1.97$^{+0.1}_{-0.05}$    & 0.392$\pm$0.002                   & 1855.5/984     \\
\hline
\end{tabular}
\end{center}
\begin{minipage}{\textwidth}
Notes: Model is abbreviated to \textsc{xspec} syntax: TBABS -
absorption components for both Galactic and external absorption;
DISKPBB - $p$-free disc. Specific notes: $^a$External absorption
column ($\times$ 10$^{22}$ cm$^{-2}$) left free during fitting,
Galactic columns listed in Table \ref{tab:sample}; $^b$ inner-disc
temperature (keV); $^c$$p$ value, where disc temperature scales as $r^{-p}$
and $r$ is the radius.
\end{minipage}
\end{table*}


\subsection{More physical models: slim disc}

The investigations and results described above have focused on the
application of simple phenomenological models used to characterise the
shape of the observed spectra of these sources. This work has revealed
the presence of both a soft excess and a break above $\sim$
2~keV. These studies have also shown the limitations of these models
at this level of data quality, especially the MCD fits. Disc spectra
should be more complex than this simple sum-of-blackbodies
approach. Firstly, relativistic effects broaden the spectrum, so it is
less strongly peaked than an MCD (Cunningham 1975; Ebisawa et
al. 2001; Li et al. 2005). Secondly, the intrinsic spectrum from
material at a given radius is not a true blackbody as there is not
enough absorption opacity to thermalise the emission at all energies
(Davis et al 2005). This can be approximated by a colour temperature
correction (Shimura \& Takahara 1995), but this is only an
approximation and the best current models of disc spectra show weak
atomic features in the continuum emission. Hence, even relativistic,
colour corrected blackbody discs are not an accurate description of
the best theoretical disc spectra over the 0.3-10~keV bandpass, though
they are normally an excellent fit to the 3-20~keV energy range (Done
\& Davis 2008). This can be seen in real data from disc dominated BHB
systems e.g. the 0.3-10~keV spectrum from LMC X--3 is better fit by
full disc models than by {\tt diskbb} (Davis, Done \& Blaes 2006).
However, only a few ULX fit well even to these full disc models
(Hui \& Krolik 2008).

\begin{figure}
\begin{minipage}{85mm}
\begin{center}
\leavevmode
\epsfxsize=8cm \rotatebox{0}{\epsfbox{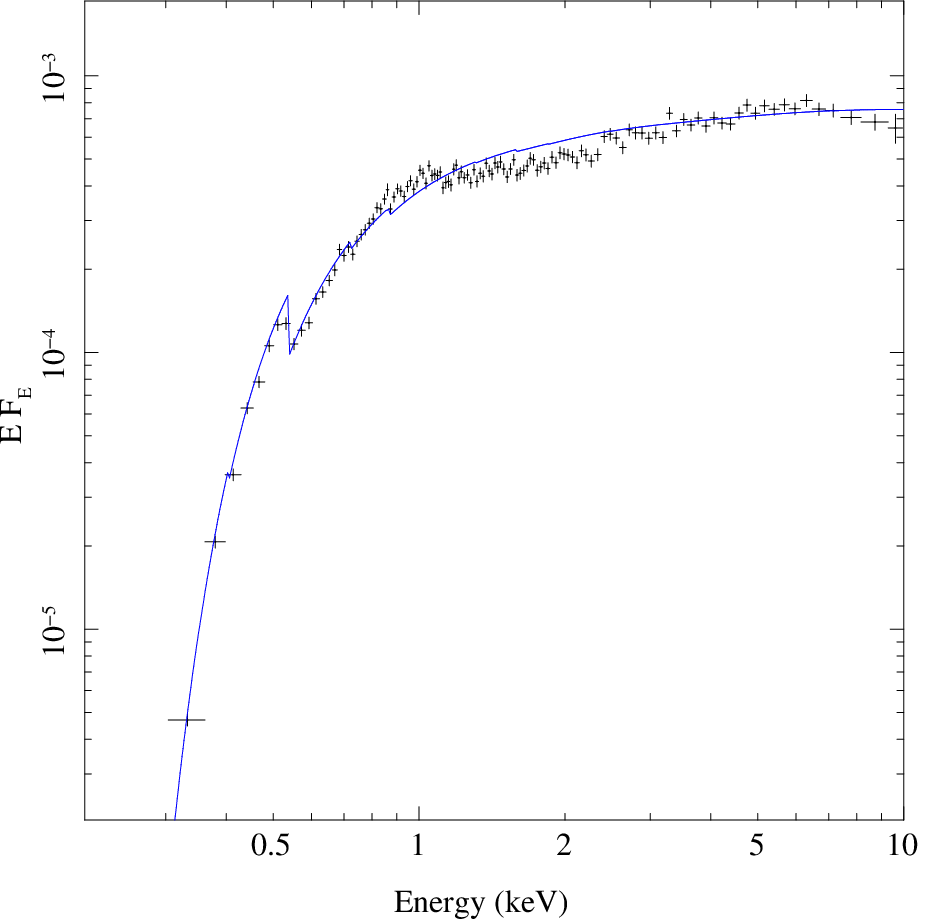}} 
\end{center}
\caption{{\it XMM-Newton} data of NGC 1313 X-1, deconvolved with the
`$p$-free' model. The data is plotted as per the previous figure, 
and binned to to a minimum of $20\sigma$ or 20 channels for clarity.
This model is used as a simplified description of the theoretically predicted `slim disc'
model. Although the $p$ value ($p$ $\sim$ 0.5) for this data could be
considered in support of a slim disc, the best fit provides an
unrealistic disc temperature ($kT_{\rm in}$ $\sim$ 7.9 keV).}
\label{fig:diskpbb}
\end{minipage}
\end{figure}

These models also assume that advection of radiation is not important,
yet in ULXs we may be observing a super-Eddington accretion flow. The
disc becomes so optically thick at these extreme mass accretion rates
that the energy released in the mid-plane of the disc does not have
time to diffuse to the photosphere. Instead, the photons are carried
along with the inflowing material, being advected radially rather than
being radiated vertically (Abramowicz et al. 1988). Such radiatively
inefficient, optically thick, advection dominated disc solutions
(termed `slim discs') are physically very different to the optically
thin, advection dominated flows (ADAFs) of Narayan \& Yi (1995), as
the energy is advected in photons rather than in protons. The
additional cooling from advected photons is most important in the
inner regions of the disc, so the expected luminosity from each disc
radius of the slim disc is progressively lower than that of a standard
accretion disc at the same mass accretion rate (Abramowicz et
al. 1988). Thus these models predict that the slim disc spectra are
less sharply curved than that of a standard disc, so these generally
give a better fit to the data. Additionally, the slim discs may extend
down to smaller radii than the classic last stable orbit due to the
non-negligible pressure support, so their spectra can include higher
temperature components than expected from a standard disc (Abramowicz
et al. 1988, Watarai et al. 2000, but see also Beloborodov 1998).
Such high temperature components, if interpreted as emission from a 
standard accretion disc, would give an underestimate of the black
hole mass (e.g. Makishima et al.  2000).

To take these physical differences into account we replace the
standard MCD model with a modified version, the `$p$-free' disc
model. This allows the disc temperature to scale as $r^{-p}$, where
$r$ is the radius and $p$ is a free parameter. Standard (MCD) discs
have $p$ fixed at 0.75, but increasing amounts of advection can be
modeled by decreasing $p$, with fully advective discs having $p$
$\simeq$ 0.5 (e.g Watarai et al. 2001, Vierdayanti et al. 2006,
Miyawaki et al. 2006).  Hence a $p$-free disc fitting with $p \sim
0.5$ would be indicative of the slim disc model, and also gives a
broader spectrum which can approximate the effects of the
relativistic effects and colour temperature correction. 

The $p$-free model gives a slightly better overall fit to the data
than a single power-law, with all ULXs being at least roughly
represented with this gentle spectral curvature.  Indeed, five out of
twelve ULXs have statistically acceptable fits to this model.  If we
consider the values derived from best fits to the data we find, at
first glance, that our results appear to be in good agreement with the
slim disc model, with $p$ $\sim$ 0.4 -- 0.6 (see Table
\ref{tab:diskpbb}), but on closer inspection problems begin to
emerge. Table \ref{tab:diskpbb} shows that while some of the derived
inner disc temperatures are in the 1 -- 3~keV range expected for such
super-Eddington flows onto stellar mass black holes, some are
extremely high, with four of the twelve sources providing fits above
6~keV (including two of the acceptable fits).  Such temperatures are
high even for maximally spinning black holes (Ebisawa et al. 2003),
making these fits physically unrealistic.  These high quality data
sets also clearly show that the spectral curvature is more complex in
many cases, even for this modified disc model.  One clear example of
this is NGC 1313 X-1 (one of the unphysically high temperature
spectra) where there is a marked inflection present in the spectral
data at $\sim$ 2 keV which cannot be matched in the $p$-free models
(Figure~\ref{fig:diskpbb}; this is also obvious in the spectra of
several other ULXs, see Figure~\ref{fig:montage}).  So, although a
$p$-free model with slim disc characteristics (or indeed a full
standard disc model: Hui \& Krolik 2008) cannot be ruled out in the
minority of cases, it does not seem to be a good explanation for ULXs
as a class. The two component phenomenological model (MCD plus
power-law) is clearly a better statistical description of the majority
of ULX spectra, though the presence of a high energy break clearly
also shows it limitations. As a next step we replace the power-law
with a Comptonisation model, to explore the nature of the high energy
break.

\subsection{Comptonisation Models} %
\label{subsection:compton}

We have demonstrated that the highest quality ULX data indicates many
of these objects are in a new, ultraluminous accretion state.  The
next, obvious question is: what are the physics of the accretion flow
producing this shape of spectrum?  In order to further explore the
nature of these systems, we now replace the power-law continuum with
more physically realistic Comptonisation models. We use an alternative
disc model, DISKPN (Gierlinski et al. 1999), which incorporates an
approximate stress-free inner boundary condition as opposed to the
continuous stress assumed in DISKBB.  The resulting spectra differ by
less than 5 \% for the same temperature, but we use DISKPN so as to be
able to directly compare our results with those of SRW06.

\renewcommand{\baselinestretch}{1.5} 
\begin{table*}
\leavevmode
\caption{Application of Comptonisation models: DISKPN+COMPTT spectral fits}
\label{tab:comptt}
\begin{center}
\begin{tabular}{l c c c c c c c}
\hline
Source               &  \multicolumn{4}{c}{TBABS*TBABS*(DISKPN+COMPTT)}                                 & \multicolumn{1}{c}{$\chi$$^2$/DoF} & $\Delta$$\chi$$^2$ $^e$ & 1-P({\it F}-test)$^f$ \\
                             & \multicolumn{1}{c}{$N_{\rm H}$$^a$} & \multicolumn{1}{c}{$T_{\rm max}$$^b$} & \multicolumn{1}{c}{$kT_{e}$$^c$}       & \multicolumn{1}{c}{$\tau$$^d$} &                &   \\
\hline
NGC 55 ULX    &  0.235$\pm$0.004                  & 0.221$^{+0.010}_{-0.006}$ & 0.83$^{+0.05}_{-0.04}$  & 9.9$\pm$0.4              & 986.3/876       & 252.6  &  $>$ 99 \\ 
M33 X-8            &  0.041$\pm$0.003                  & 0.87$^{+0.04}_{-0.2}$           & 1.39$^{+0.08}_{-0.03}$  & 80$^{+100}_{-30}$  & 1204.4/1244   & 42.2    &  $>$ 99    \\ 
NGC 1313 X-1 &  0.21$\pm$0.01                       & 0.23$\pm$0.01                      & 2.1$\pm$0.1                      & 8.5$^{+0.6}_{-0.5}$  & 1705.7/1608   & 192.0  &   $>$ 99 \\
NGC 1313 X-2 &  0.195$^{+0.004}_{-0.005}$ & 0.7$^{+0.2}_{-0.1}$               & 1.51$\pm$0.02                 & 15.3$^{+0.4}_{-1}$   & 1613.2/1592   & 281.6  &   $>$ 99  \\ 
IC 342 X-1        &  0.58$^{+0.10}_{-0.1}$           & 0.30$^{+0.2}_{-0.08}$          & 2.78$^{+8}_{-0.6}$          & 7$^{+3}_{-7}$            & 515.5/508        & 4.4      &  87.5  \\
NGC 2403 X-1 &  0.18$^{+0.06}_{-0.01}$        & 0.27$^{+0.1}_{-0.05}$          & 0.98$\pm$0.04                 & 12.5$\pm$0.7            & 844.7/835        & 131.4 &  $>$ 99  \\ 
Ho II X-1            &  0.033$\pm$0.002                  & 0.23$^{+0.02}_{-0.01}$        & 2.12$^{+0.1}_{-0.10}$    & 5.5$\pm$0.2               & 1375.4/1360   &  44.2   &   $>$ 99  \\ 
M81 X-6            &  0.181$\pm$0.006                  & 0.7$^{+0.1}_{-0.2}$               & 1.15$^{+0.02}_{-0.01}$  & 31$^{+4}_{-3}$          & 1082.7/982     &  56.8   &   $>$ 99  \\ 
Ho IX X-1          &  0.099$\pm$0.006                  & 0.26$\pm$0.02                       & 2.27$^{+0.1}_{-0.09}$    & 9.5$\pm$0.5               & 2284.2/2109  & 126.4  &  $>$ 99  \\ 
NGC 4559 X-1 &  0.13$^{+0.03}_{-0.02   }$     & 0.16$\pm$0.02                       & 1.8$^{+0.6}_{-0.3}$         & 7.4$^{+0.7}_{-1}$     & 513.1/585       & 32.4    &  $>$ 99  \\ 
NGC 5204 X-1 &  0.035$^{+0.007}_{-0.009}$ & 0.26$\pm$0.03                       & 2.2$^{+2}_{-0.4}$            & 6 $^{+2}_{-3}$           & 886.2/865        & 3.9     &  87.5 \\
NGC 5408 X-1 &  0.029$\pm$0.005                  & 0.170$^{+0.006}_{-0.007}$ & 1.5$^{+0.3}_{-0.2}$         & 6.5 $^{+0.8}_{-0.9}$ & 1240.9/982      & 41.1   &   $>$ 99  \\ 
\hline
\end{tabular}
\end{center}
\begin{minipage}{\textwidth}
Notes: models are abbreviated to \textsc{xspec} syntax: DISKPN - accretion disc model, COMPTT - Comptonisation model. Specific notes: $^a$External absorption column in units of 10$^{22}$ atoms cm$^{-2}$, $^b$maximum temperature in the accretion disc (keV), $^c$plasma temperature in the Comptonising corona, $^d$optical depth of corona, $^{e,f}$$\chi$$^2$ improvement , and statistical probability (in percent) of the fit improvement over a hot optically thin corona with $kT_e = 50$ keV fixed.
\end{minipage}
\end{table*}

\subsubsection{DISKPN+COMPTT}
\label{subsubsection:comptt}

We initially use COMPTT (Titarchuk 1994) to model the coronal
emission. This is an analytic approximation to non-relativistic
thermal Comptonisation which assumes that the seed photons for
Comptonisation have a Wien spectrum.  We tie the temperature of these
seed photons to the temperature of the accretion disc ($T_{\rm
max}$). In each case the redshift is fixed at zero due to the
proximity of these systems, whilst the optical depth and plasma
temperature are free to vary.  We show the results of our spectral
fitting in Table~\ref{tab:comptt}.

\begin{figure}
\begin{minipage}{85mm}
\begin{center}
\leavevmode
\epsfxsize=6.1cm \rotatebox{-90}{\epsfbox{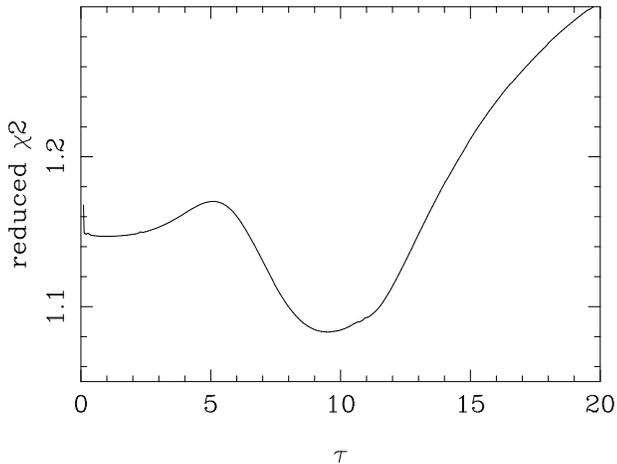}} 
\end{center}
\caption{Variation in $\chi$$^2$ over a range of optical depths derived from the fitting of an absorbed DISKPN plus COMPTT to the data from Ho IX X-1. A local minimum is observed at lower optical depths but this clearly shows that the global minimum occurs at $\tau$ $\sim$ 9.5,  an optically thick solution.}
\label{fig:tau}
\end{minipage}
\end{figure}

We find in each case that there is a local minimum in $\chi$$^2$ space
which corresponds to a hot, optically thin Comptonising corona
($kT_{\rm e}$ $\sim$ 50 keV, $\tau$ $\la$ 1), similar in nature to
that seen in Galactic BHBs in classic accretion states.  However, the
global minimum in $\chi$$^2$ occurs for a fit that describes a cool,
optically thick corona ($kT_{\rm e}$ $\sim$ 1 - 3 keV and $\tau$
$\sim$ 6 - 80).  This is a very different scenario to those of the
standard black hole accretion states.  We illustrate this in Figure
\ref{fig:tau}, which shows $\chi$$^2$ versus $\tau$ for Ho IX
X-1. Here we find that a local minimum is observed at low optical depths,
but a clear global minimum occurs at $\tau \sim$ 9.5, an optically
thick solution.

To quantify this improvement, we have compared the fits achieved in
each scenario and find that in ten of the twelve 
spectra the improvement in
$\chi$$^2$ is highly significant ($\Delta$$\chi$$^2$ $>$ 30) (see
Table~\ref{tab:comptt}), while the other two still show
$\Delta$$\chi$$^2$ $>$ 3.9 i.e. the break is detected at 90\%
significance. To test this further we fix the temperature
of the corona at 50 keV (all hot corona local minima fits lie at
approximately this temperature within errors). We compare the
resultant fits to those given in Table \ref{tab:comptt} using the
F-test, finding that in ten out of twelve cases in our sample shows a
$>$ 99 \% statistical probability of improvement when applying a cool
optically thick corona over a hot optically thin one.  The possible
presence of such an extraordinary corona is therefore the first major
clue as to the origin in the physical differences between
sub-Eddington accretion states, and the ultraluminous state.

\renewcommand{\baselinestretch}{1.5} 
\begin{table*}
\leavevmode
\caption{Application of Comptonisation models: DISKPN+EQPAIR spectral fits}
\label{tab:eqpair}
\begin{center}
\begin{tabular}{l c c c c c}
\hline
Source               &  \multicolumn{4}{c}{TBABS*TBABS*(DISKPN+EQPAIR)}                               & \multicolumn{1}{c}{$\chi$$^2$/DoF} \\
                             & \multicolumn{1}{c}{$N_{\rm H}$$^a$} & \multicolumn{1}{c}{$T_{\rm max}$$^b$} & \multicolumn{1}{c}{$l_{\rm h}/l_{\rm s}$$^c$}       & \multicolumn{1}{c}{$\tau$$^d$} & \\
\hline
NGC 55 ULX    &  0.250$^{+0.01}_{-0.004}$   & 0.253$^{+0.01}_{-0.005}$   & 1.24$\pm$0.03                    & 26.9$^{+0.7}_{-2}$               & 990.2/876     \\ 
M33 X-8            &  0.036$^{+0.004}_{-0.005}$ & 0.63$\pm$0.02                       & 0.70$^{+0.03}_{-0.02}$     & 13.2$^{+0.5}_{-0.8}$            & 1204.0/1244  \\ 
NGC 1313 X-1 &  0.210$^{+0.005}_{-0.01}$   & 0.271$^{+0.004}_{-0.010}$ & 4.02$\pm$0.07                    & 17.9$^{+0.5}_{-0.6}$            & 1711.6/1608    \\ 
NGC 1313 X-2 &  0.204$\pm$0.01                    & 0.37$^{+0.02}_{-0.04}$        & 2.28$^{+0.05}_{-0.06}$      & 17.7$^{+0.6}_{-0.8}$           & 1601.8/1592    \\ 
IC 342 X-1         &  0.64$^{+0.08}_{-0.07}$       & 0.32$^{+0.02}_{-0.05}$        & 3.2$^{+0.5}_{-0.1}$             & 11$^{+3}_{-0.3}$                  & 516.1/508         \\ 
NGC 2403 X-1 &  0.24 $^{+0.04}_{-0.02}$       & 0.32$^{+0.04}_{-0.05}$        & 1.87$^{+0.05}_{-0.07}$      & 27.6$\pm$2                          & 849.3/835       \\ 
Ho II X-1            &  0.050$^{+0.002}_{-0.001}$ & 0.267$^{+0.007}_{-0.003}$ & 1.011$^{+0.009}_{-0.01}$ & 7.00$\pm$0.02                     & 1391.2/1360  \\ 
M81 X-6            &  0.20$^{+0.02}_{-0.03}$        & 0.30$^{+0.1}_{-0.02}$          & 2.6$^{+0.1}_{-0.6}$             & 25$^{+3}_{-1}$                     & 1085.2/982         \\ 
Ho IX X-1          &  0.109$^{+0.005}_{-0.004}$ & 0.312$^{+0.01}_{-0.006}$   & 4.99$\pm$0.08                    & 18.5$^{+0.4}_{-0.3}$            & 2293.1/2109       \\ 
NGC 4559 X-1 &  0.14$^{+0.01}_{-0.02}$        & 0.177$\pm$0.009                  & 2.2$\pm$0.1                         & 14$\pm$2                              & 513.8/585          \\ 
NGC 5204 X-1 &  0.047$\pm$0.003                  & 0.314$\pm$0.01                    & 1.35$^{+0.07}_{-0.05}$      & 10$^{+1}_{-2}$                     & 894.3/865          \\ 
NGC 5408 X-1 &  0.033$^{+0.002}_{-0.001}$ & 0.184$^{+0.003}_{-0.002}$ & 1.140$^{+0.02}_{-0.009}$ & 12.9$^{+0.4}_{-0.6}$           & 1244.2/982       \\ 
\hline
\end{tabular}
\end{center}
\begin{minipage}{\textwidth}
Notes: models are abbreviated to \textsc{xspec} syntax: DISKPN - accretion disc model, EQPAIR - Comptonisation model. Specific notes: $^a$External absorption column in units of 10$^{22}$ atoms cm$^{2}$, $^b$maximum temperature in the accretion disc (keV), $^c$ratio between the compactness of electron and the compactness of seed photon distribution, $^d$optical depth.
\end{minipage}
\end{table*}

A second characteristic found in fitting this model to our ULX data is
that the disc temperatures are generally cool.  In fact, Table
\ref{tab:comptt} shows that 9/12 disc temperatures reside (on face
value) in the IMBH range ($T_{\rm max} <$ 0.5 keV), while the
remaining three objects have hotter discs that might imply stellar
mass objects.  Similar results -- a combination of a cool disc, and an
optically thick cool Comptonising corona -- have been found in
previous studies of individual sources (e.g. Ho IX X-1 by Dewangan et
al. 2006; Ho II X-1 in Goad et al. 2006; also SRW06).  However, in
interpreting the cool disc component we must consider the implications
of the presence of an optically thick corona.  It is likely that such
a medium would mask the inner-most radii of the disc.  It is also
likely that material and power may be drawn from the disc to feed the
corona. A combination of these factors could greatly impact the
observed temperature of the accretion disc, invalidating any
conclusions drawn on this basis.  We return to this point later.

\subsubsection{DISKPN+EQPAIR}
\label{subsubsection:eqpair}

We now apply an alternative, more physically self-consistent
Comptonisation model to test our provisional result from the
DISKPN$+$COMPTT model further. Here we use EQPAIR (Coppi 1999), a
model which allows both thermal and non-thermal electrons, and
calculates the resulting spectrum without assuming that the electrons
are non relativistic, and where the seed photons can have a disc or
blackbody spectrum. We choose to use thermal electrons plus disc
emission to allow for comparison to COMPTT, and again tie the
temperature of the seed photons to that of the inner accretion disc.

Comparing the fits to the data for DISKPN+COMPTT (Table
\ref{tab:comptt}) with DISKPN+EQPAIR (Table \ref{tab:eqpair}), we find
that similar $\chi^2$ values are achieved in all cases. Results once
again indicate that the global fitting minima are achieved by a
relatively cool accretion disc ( 0.18 $<$ $T_{\rm max}$ $<$ 0.63 keV)
with a cool optically thick corona ($\tau \sim$ 7.0--27.6), thick
enough to potentially hide the inner disc.

\begin{figure}
\begin{minipage}{85mm}
\begin{center}
\leavevmode
\epsfxsize=6.1cm \rotatebox{-90}{\epsfbox{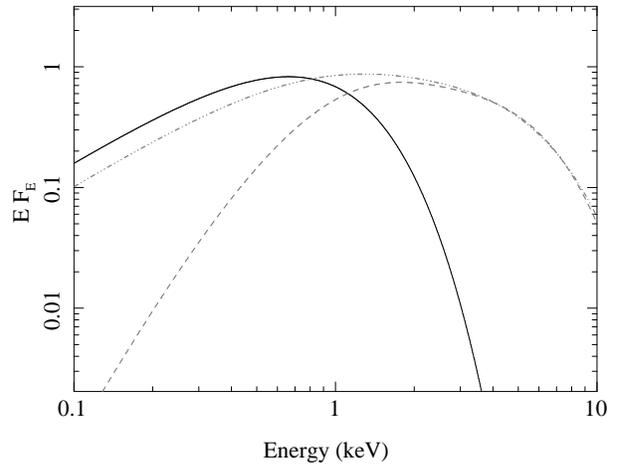}} 
\end{center}
\caption{Comparison of disc plus Comptonisation models used within our analysis. An average cool disc is plotted in black with coronal components representative of those found in our fits shown in grey (dashed line -- COMPTT; dashed \& dotted line -- EQPAIR). Although the Comptonisation components have different spectral structure at lower energies, it is clear that the same cool optically thick corona is achieved at higher energies of the {\it XMM-Newton} band pass. The differences that are evident in the lower part of our energy range are accounted for by the other components in our model. }
\label{fig:compton}
\end{minipage}
\end{figure}

Figure \ref{fig:compton} illustrates the similarity in results between
the two Comptonisation models.  Here we plot typical values from our
fits for both Comptonisation components in grey (COMPTT -- dashed
line; EQPAIR -- dashed \& dotted line) along with an average disc
shown in black. We can clearly see that each of these models exhibits
similar curvature at higher energies. The difference in the spectrum
of these coronal components lies at lower energies where subtle
variations in the disc and absorption components can compensate for
this. Each of these models can therefore explain the curvature
observed at both higher and lower energies within our band-pass. The
soft excess at low energies is modelled well by a cool accretion disc,
whilst the optically thick corona causes the downturn at high energies
(irrespective of model choice).

When these results are compared to the findings of SRW06, we find that
we achieve similar fits to the data\footnote{We have five data sets in
common with SRW06, and a further four sources with improved
data.}. Their results also indicate the presence of a cool accretion
disc (0.08 $< kT_{\rm max} <$ 0.29 keV) in all cases, whilst an
optically thick corona is observed in almost all sources ($\tau$
ranges from 0.2--33).  It is therefore clear that these appear to be
the physical characteristics of ULXs, and hence the ultraluminous
state.  This clearly demonstrates that we appear to be observing a
radically different accretion flow to the classic accretion states
viewed in Galactic BHBs.

We should note, however, that there are at least two Galactic black
hole candidates that exhibit similar (albeit less extreme) spectral
traits when modelled similarly -- an unusually cool accretion disc,
and a cool optically thick corona -- to the ULXs observed within our
sample (when band-pass is considered). These objects are GRS 1915+105
and XTE J1550-564, and in each case these objects are thought to be
accreting at, or above, the Eddington limit when displaying such
traits (Kubota \& Done 2004; Middleton et al. 2006; Ueda et al. 2009).  This is further
evidence to support the assertion that the ultraluminous state
represents a super-Eddington accretion flow.

\subsection{Energetic disc-corona coupling} %
\label{subsection:chris}

\begin{figure}
\begin{minipage}{85mm}
\begin{center}
\leavevmode
\epsfxsize=7.9cm \rotatebox{270}{\epsfbox{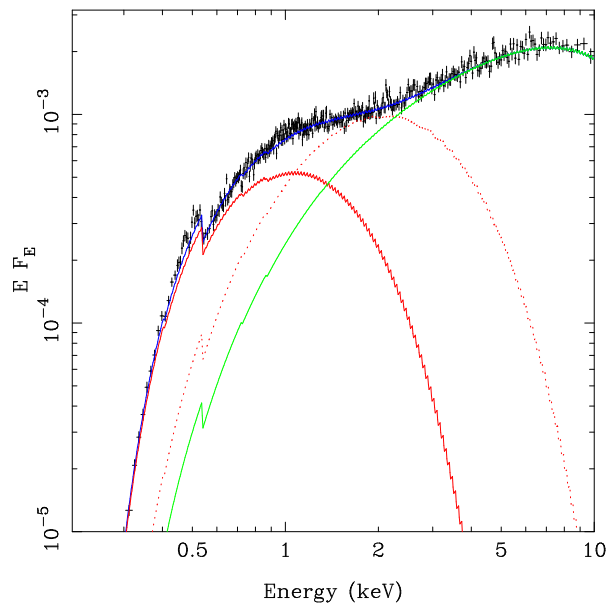}} 
\end{center}
\caption{{\it XMM-Newton} EPIC pn data from Ho IX X-1, fit with an absorbed ultraluminous model (DKBBFTH, in blue).  We plot only the pn data, shown in black, which we rebin to a minimum of $15\sigma$ significance or 15 channels for clarity.  We also plot the various components of the accretion system described by the of the model, once again disc components are plotted in red whilst the corona is in green. The visible regions of the outer disc and the optically thick corona are plotted with a solid line, whilst the masked emission from the cooled, energetically-coupled inner disc is represented by a dotted line. This shows that the curvature at lower energies is due to emission from the outer accretion disc (we are not able to directly observe emission from the innermost regions), and the break or turnover at higher energies is caused by the emission from the optically thick corona.}
\label{fig:chris}
\end{minipage}
\end{figure}

\renewcommand{\baselinestretch}{1.5} 
\begin{table*}
\leavevmode
\caption{The ultraluminous model: DKBBFTH}
\label{tab:chris}
\begin{center}
\begin{tabular}{l c c c c c c c c}
\hline
Source               &  \multicolumn{7}{c}{TBABS*TBABS*(DKBBFTH)}                               & \multicolumn{1}{c}{$\chi$$^2$/DoF} \\
                            & \multicolumn{1}{c}{$N_{\rm H}$$^a$} & \multicolumn{1}{c}{$kT_{\rm disc}$$^b$} & \multicolumn{1}{c}{$R_{\rm c}/R_{\rm in}$$^c$}  & \multicolumn{1}{c}{$R_{\rm in}$$^d$}  & \multicolumn{1}{c}{$\Gamma$$^e$} & \multicolumn{1}{c}{$kT_{\rm e}$$^f$} & \multicolumn{1}{c}{$\tau$$^g$}  \\
\hline
NGC 55 ULX    &  0.239$^{+0.02}_{-0.006}$  	& 0.38$^{+0.02}_{-0.01}$     	& 1.5$^{+0.1}_{-0.2}$     	& 700$^{+300}_{-100}$    & 2.0$^{+0.3}_{-0.5}$         & 0.79$^{+0.1}_{-0.08}$  & 10.6$^{+0.3}_{-0.5}$    & 989.1/876  \\ 
M33 X-8       	&  0.036$^{+0.004}_{-0.006}$ 	&  1.03$\pm$0.04             		& 2.7$^{+1}_{-0.7}$       	& 65$^{+10}_{-6}$     & 3.6$^{+0.1}_{-1}$           & 6$^{+30}_{-4}$         & 1.5$^{+0.2}_{-1}$       & 1201.9/1244     \\ 
NGC 1313 X-1  &  0.211$^{+0.004}_{-0.005}$ 	& 0.60$^{+0.03}_{-0.04}$     	& 2.82$^{+0.10}_{-0.07}$ & 460$^{+40}_{-30}$  & 1.66$^{+0.03}_{-0.02}$      & 2.19$^{+0.2}_{-0.09}$  & 7.79$^{+0.03}_{-0.02}$  & 1708.5/1608     \\ 
NGC 1313 X-2  &  0.21$^{+0.01}_{-0.02}$   	 & 1.18$^{+0.05}_{-0.1}$      	& 4.3$^{+1}_{-0.5}$       	& 130$^{+80}_{-30}$    & 2.5$^{+0.1}_{-0.2}$         & 3$^{+6}_{-1}$          & 3.5$^{+0.1}_{-0.2}$     & 1594.8/1592     \\ 
IC 342 X-1    	&  0.55$^{+0.08}_{-0.03}$    	& 1.0$\pm$0.2                		& 2.8$^{+3}_{-0.8}$       	& 160$^{+400}_{-80}$    & 1.6$^{+0.2}_{-0.5}$         & 2.4$^{+7}_{-0.7}$      & 7.9$^{+0.2}_{-0.5}$     & 515.7/508    \\ 
NGC 2403 X-1  &  0.234$^{+0.009}_{-0.03}$  	& 0.85$^{+0.08}_{-0.1}$     	 & 2.4$^{+0.4}_{-0.6}$     	& 170$^{+40}_{-30}$    & 1.9$^{+0.2}_{-0.4}$         & 1.003$^{+10}_{-0.003}$ & 9.7$^{+0.2}_{-0.4}$     & 846.9/835        \\ 
Ho II X-1     	&  0.079$^{+0.008}_{-0.007}$ 	& 0.30$^{+0.02}_{-0.01}$     	& 5.6$^{+1.0}_{-0.1}$     	& 3900$\pm$100   & 2.53$^{+0.02}_{-0.01}$      & 13.2$^{+1}_{-0.6}$     & 1.43$^{+0.02}_{-0.01}$  & 1399.4/1360      \\ 
M81 X-6       	&  0.19$\pm$0.02             		& 0.98$^{+0.08}_{-0.05}$     	& 3.3$\pm$0.2             	& 140$\pm$40  & 2.15$^{+0.2}_{-0.07}$       & 1.39$^{+0.07}_{-0.2}$  & 6.87$^{+0.2}_{-0.07}$   & 1080.7/982     \\ 
Ho IX X-1     	&  0.121$^{+0.008}_{-0.005}$ 	& 1.01$^{+0.03}_{-0.04}$     	& 4.0$^{+0.2}_{-0.1}$     	& 220$^{+50}_{-30}$    & 1.58$\pm$0.03               & 2.5$\pm$0.2            & 7.92$^{+0.03}_{-0.03}$  & 2286.9/2109        \\ 
NGC 4559 X-1  &  0.138$^{+0.008}_{-0.01}$  	& 0.31$^{+0.06}_{-0.04}$     	& 2.0$^{+0.2}_{-0.1}$     	& 3000$^{+3000}_{-2000}$   & 1.95$^{+0.09}_{-0.10}$      & 1.9$^{+0.6}_{-0.3}$    & 6.73$^{+0.09}_{-0.10}$  & 513.8/585        \\ 
NGC 5204 X-1  &  0.036$\pm$0.007           	& 0.54$\pm$0.02              		& 1.8$^{+0.2}_{-0.1}$     	& 610$^{+100}_{-90}$    & 2.0$\pm$0.2                 & 1.9$\pm$0.4            & 6.5$^{+0.2}_{-0.2}$     & 889.1/865        \\ 
NGC 5408 X-1  &  0.029$^{+0.007}_{-0.004}$ & 0.255 $^{+0.006}_{-0.005}$ 	& 1.47$^{+0.01}_{-0.05}$ & 3000$^{+200}_{-40}$   & 2.31$^{+0.04}_{-0.03}$      & 1.6$^{+0.2}_{-0.1}$    & 5.80$^{+0.04}_{-0.03}$  & 1246.5/982    \\ 
\hline
\end{tabular}
\end{center}
\begin{minipage}{\textwidth}
Notes: models are abbreviated to \textsc{xspec} syntax: DKBBFTH - energetically coupled disc - Comptonised corona model. Specific notes:  $^a$External absorption column in units of 10$^{22}$ atoms cm$^{2}$, $^b$un-Comptonised disc  temperature (keV), $^c$external radius of corona, $^d$inner radius of the accretion disc, $^e$photon index,  $^f$temperature of the Comptonising corona, $^g$$\tau$ is not a fit parameter of the model, but is derived from $\Gamma$ and $kT_e$.
\end{minipage}
\end{table*}
\renewcommand{\baselinestretch}{1.0} 

We now move on to the final stage in our analysis to further explore
the results provided by Comptonisation models. The models above
implicitly assume that we can still observe the disc down to its
innermost stable orbit. This assumption requires that the optically
thick corona does not intercept our line of sight to the inner disc
and that the underlying disc spectrum is independent of that of the
corona (see Kubota \& Done 2004). Yet both assumptions are probably
flawed.  An optically thick corona could easily mask the innermost
regions (depending on its geometry: Kubota \& Done 2004).  Secondly,
we must consider the energetics of both components.  Both the disc and
the corona must be ultimately powered by gravitational energy
release. If we are observing a powerful corona, this implies less
energy is available for heating the disc (Svensson \& Zdziarski 1994). The
only code currently available that enables us to explore the effect of
relaxing these assumptions is DKBBFTH (Done \& Kubota 2006). This was
designed to model the extreme very high state spectra seen in BHBs
such as XTE J1550-564 and GRS 1915+105. It incorporates the energetic
disc-corona coupling model of Svensson \& Zdziarski (1994), which
assumes that the corona extends over the inner disc from $R_{\rm in}$
to $R_{\rm T}$, taking a fraction $f$ of the gravitational
energy available at these radii. Only the remaining fraction $(1-f)$ is
available to power the inner disc emission, so the inner disc is
cooler, but more importantly, less luminous than it would be if the
corona were not present. This distorted inner disc spectrum is the
source of seed photons for the Comptonisation (Done \& Kubota 2006).

This model has five free parameters (only one more than in the purely
phenomenological MCD plus power law model and the same as for all the
disc plus Comptonisation models).  The first two are the temperature of
the Comptonising plasma, $kT_e$, and its optical depth (parameterised
by the asymptotic spectral index, $\Gamma$).  The disc is assumed to
extend from some inner radius, $R_{in}$, which is given by the overall
normalisation of the model, and with a temperature distribution {\em
in the limit of no corona being present} of $T(R)_0=T_{in}
(R/R_{in})^{-3/4}$ i.e. as in the {\tt diskbb} model. However, the model
assumes that the corona extends homogeneously (constant temperature
and optical depth) in a slab over the disc, from its inner radius to
$R_{\rm T}$ (in units of the Schwartzschild radius, $R_{\rm s} =
2GM/c^2$). For all radii between this and $R_{in}$, the calculated
disc temperature is $T(R)=T(R)_{0} (1-f)^{1/4}$ where $f$ is the
fraction of power dissipated in the corona (assumed constant with
radius) which is self consistently calculated via iteration from the 
coronal spectral parameters (Done \& Kubota 2006). The resulting 
disc luminosity underneath the corona is reduced by a factor $(1-f)$, 
but only a fraction $e^{-\tau}$ of this is seen directly, with the remainder 
being Compton scattered by the corona  (Done \& Kubota 2006).

Despite being more physically constrained, the model gives an
equivalently good fit to the data as the disc plus Comptonisation
models discussed above. Figure \ref{fig:chris} shows this fit for Ho
IX X-1, with the outer (un-Comptonised) disc emission (red solid line)
dominating at soft energies. The inner disc emission (red dotted line)
is much lower than in standard disc models as much of the energy is
powering the corona. These form the seed photons for the Compton
scattering, but since this scattering is in an optically thick corona,
these photons are {\em not} seen as almost all of them are upscattered
to form the Comptonised spectrum (solid green line).

Figure \ref{fig:montage} shows the model fit to all our sources, with
the pn data corrected for absorption to show the intrinsic spectral
shape.  This clearly shows the difficulties in interpreting the
parameters from simple, phenomenological spectral fits. Again, using
Ho IX X-1 as an example, the data have an inflection at soft energies
that characterises the soft excess. The characteristic energy of this
feature is interpreted as the peak of the disc emission in the MCD
plus power-law model (giving a very low temperature: hence IMBHs),
while the high energy break forms the peak of the disc temperature in
single component disc models (giving a very high temperature: too
extreme even for stellar mass black holes).  In these coupled
disc-corona models, the `true' disc temperature is not given by either
of these observed features!  The high energy break is from the very
low plasma temperature of the Comptonising region, while the low
energy inflection occurs when the outer, un-Comptonised disc emission
starts to dominate the spectrum. The inferred intrinsic inner disc
temperatures, recovered from the assumption of disc-corona energy
partition, range from 0.3 $<$ $kT_{\rm disc}$ $<$ 1.2 keV with 8/12
giving fits where $kT_{\rm disc}$ $>$ 0.5 keV, which lies in the
stellar mass black hole regime. This can be seen explicitly by
converting the inferred inner radius into black hole mass assuming
that the disc extends down to the last stable orbit around a
Schwarzschild black hole at $R_{in}=6GM/c^2=8.9M$. All these `hot'
ULXs ($kT_{\rm disc}$ $>$ 0.5 keV) have inferred black hole masses
$<100~M_\odot$, consistent with stellar mass black holes, and
consequent derived $L_X/L_{Edd}\ga 1$.

However, a third of these source spectra are best fit by $kT_{\rm
disc}$ $<$ 0.5 keV, hence giving black hole masses in the range
80-430~$M_\odot$ and {\em sub}-Eddington accretion rates assuming that
the disc extends down to the last stable orbit. We propose instead
that this very cool temperature and hence large radius is {\em not}
associated with the direct disc emission but instead arises as the
accretion rate increases beyond that of standard super-Eddington
accretion. In a super-critical accretion regime, the inner regions of
the disc and corona may be blown off in the form of a wind, forming an
optically thick photosphere out to large radii (e.g. Poutanen et
al. 2007). We return to this point in the next section.

\section{Discussion: ultraluminous X-ray sources, super-Eddington accretion and the ultraluminous state}
\label{section:discussion}

We have examined twelve of the highest quality ULX data sets currently
available in the public archives of the {\it XMM-Newton\/}
telescope. Initial characterisation of the spectrum of these sources
reveals the presence of both a soft excess and a break at higher
energies (within the {\it XMM-Newton\/} band pass).  The existence of
curvature at lower energies is apparent in many of the standard BHB
accretion states, and is regularly fit by an accretion disc (Done,
Gierlinski \& Kubota 2007). It is this standard practice which led to
the suggestion that these objects were intermediate mass black holes,
due to the low temperature of the apparent disc emission.

However, the detection of a break or curvature at higher energies
($\sim$~5~keV) brings new insights. Such a feature has been observed
in ULXs previously (e.g.  Foschini et al. 2004; Feng \& Kaaret 2005;
SRW06; Miyawaki et al 2009), but here we show that it is nearly
ubiquitous in the highest quality spectral data, with $\ga$~10,000
counts. No such break at these low energies is observed in the high
energy tail of any of the standard accretion states observed in BHBs
(Remillard \& McClintock 2006; Done, Gierlinski \& Kubota
2007)\footnote{Whilst there is a break in the high energy tail of the
low/hard and very high states, it is at energies $\ga 100$ keV and
20-30~keV, respectively cf. McClintock \& Remillard (2006).} so this
challenges the basic assumption of the IMBH model, which is that we
are observing standard accretion states that are scaled with the mass
of the compact object. Thus we must consider the alternative, that we
are observing a different accretion state than those generally seen in
BHBs. A new accretion state would require us to be observing a
different mass accretion rate with respect to Eddington than those
seen in the standard states. Since these span $(\sim~10^{-7}-
1)~L_{\rm Edd}$ (McClintock \& Remillard 2006) then this is most
likely a super-Eddington state, so requires a stellar mass rather than
IMBH accretor.

We therefore suggest that a new observational state should be defined
based on the characteristic signatures of ULXs. The {\bf ultraluminous
state} is one in which we observe a new combination of observational
signatures; both a cool disc and a break or roll over at high energies
in the band pass of the {\it XMM-Newton} telescope.  We caution that
high quality data is required for a clear identification; the high
energy break in particular is difficult to identify with less than
$\sim 10,000$ counts in the X-ray spectrum.

\begin{figure*}
\begin{center}
\leavevmode
\epsfxsize=18cm \rotatebox{-90}{\epsfbox{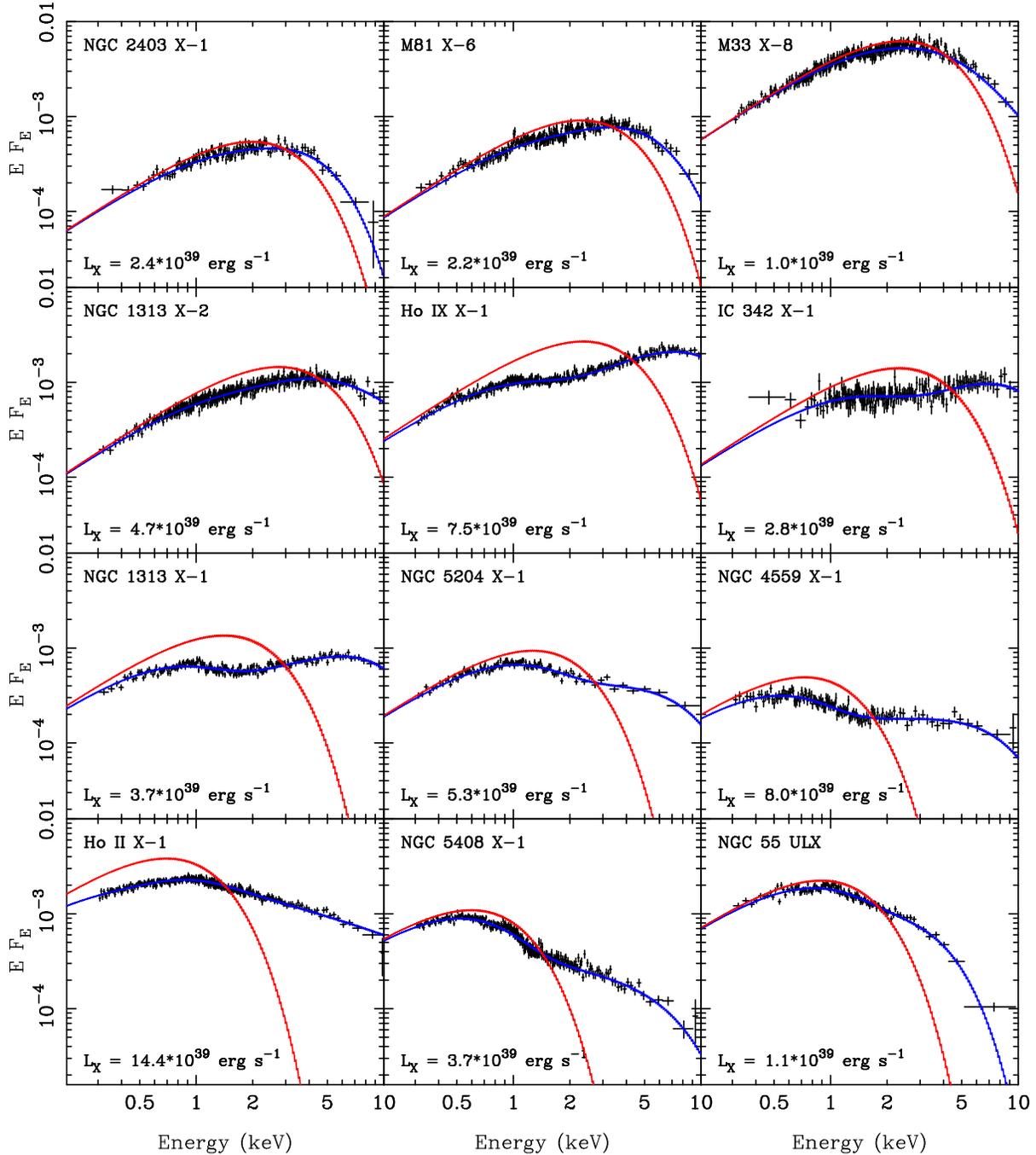}} 
\end{center}
\caption{{\it XMM-Newton} EPIC pn data (black) for all sources in our sample, absorption-corrected and deconvolved with DKBFTH (shown in blue). The `true' disc spectrum is over-plotted in red, this is the disc spectrum that would be observed in each case if the corona was removed. It is evident from these spectral plots that we are observing a variety of spectral shapes. The first four objects (NGC 2403 X-1 , M81 X-6, M33 X-8 and NGC 1313 X-2) appear very disc-like in structure and could be representative of the high or very high state. As we move further down the plots an inflection begins to emerge at $\sim$ 2 keV, signifying a break from the standard sub-Eddington accretion states, which we suggest represents a transition to a new super-Eddington accretion state. As the apparent disc temperature cools, the spectrum tips, indicating the possible presence of strong winds enveloping the inner regions of the accretion disc, leading to the most extreme cases (Ho II X-1, NGC 5408 X-1 \& NGC 55 ULX). }
\label{fig:montage}
\end{figure*}

In order to explore the physical origins of this state we applied more
physically motivated models to the data.  One theory that has been
presented to explain such an accretion state is the `slim disc' model,
where advection of radiation suppresses the emitted disc luminosity of
the innermost regions (Abramowicz et al. 1988). We apply a simplifed
model often used for such spectra, where the temperature profile in
the disc is assumed to be $T \propto$ $r^{-p}$, with $p$ a free
parameter rather than fixed at $0.75$ as for standard discs (Watarai
et al. 2001). Our data give 0.4 $\leq$ $p$ $\leq$ 0.6, similar to the
$p$ $=$ 0.5 expected for advection dominated discs, but the derived
inner disc temperatures are unrealistically high ($T_{\rm in}$ $\geq$
6 keV) for one third of our sample.  Figure \ref{fig:diskpbb}
illustrates this for NGC 1313 X-1, where $T$ $\sim$ 8~keV.  Plainly,
the disc temperature is set by the presence of the high energy break,
but such high inner disc temperatures are not expected even for
stellar mass black holes (Ebisawa et al 2003). However, we caution
against interpreting these parameters physically as the model fails to
account for the inflection that occurs at $\sim$ 2 keV. These slim
disc models cannot simultaneously produce {\em both} the soft excess
at low energies {\em and} the high energy break seen in the high
quality data used here, though this deficiency may be hidden in lower
signal-to-noise data.

In order to explore the nature of these sources in more detail, we
consider a combination of a disc plus Comptonisation due to its
successes in describing the emission of other accretion-powered BHB
systems. Two different Comptonisation models are applied to the data,
COMPTT and EQPAIR, giving two slightly different approximations to
thermal Compton up-scattering of accretion disc photons (see Figure
\ref{fig:compton}). Irrespective of which model is used, we find
evidence for a cool, optically thick corona, where the high energy
break is set by the electron temperature of this Comptonising plasma.
The parameters of this corona are rather different than anything
observed in any of the standard accretion states of BHBs. It has lower
temperature and higher optical depth than even the most extreme
optically thick corona seen in the very high state (Done \& Kubota
2006). Such material must block our view of the inner disc (Kubota \&
Done 2004), and most likely also changes the energy dissipated in the
disc (Done \& Kubota 2006).

We attempt to recover the intrinsic disc spectrum by modelling a
corona over the inner disc that both Comptonises the inner regions of
the disc and is energetically coupled to it. Both corona and disc are
ultimately powered by gravity, so increasing the power dissipated in
the corona at a given radius must mean that the disc is less luminous
than expected at that radius (Done \& Kubota 2006; Svensson \&
Zdziarski 1994). Again the parameters indicate a more extreme version
of the very high state, but unlike the phenomenological disc plus
Comptonisation models, these coupled disc-corona models allow us to
infer what the source would have looked like without any corona.

The spectral energy distributions derived on the basis of this model
can be put into a potential sequence of ULX spectra, as shown in
Figure \ref{fig:montage}. The first class are those which have spectra
which increase monotonically in $\nu f_\nu$, with maximum power output
at the energy given by the high energy break.  All these sources give
a `hot disc' ($kT_{\rm in}>1$~keV) in the canonical MCD plus power-law
fits, with the `power-law' producing the additional flux at the
softest energies. This is probably due to the MCD model providing a
poor description of the broader spectra expected from more realistic
disc models (Done \& Davis 2008; Hui \& Krolik 2008).  However, these
spectra still look fairly similar to the standard disc spectra seen in
the disc-dominated state (NGC 2403 X-1; M81 X-6 and M33 X-8).  Small
amounts of emission from a hot corona can also contribute to the
spectrum at the highest energies (see e.g. the spectral decompositions
for the most luminous states on XTE J1817-330 in Figure 4 of
Gierlinski, Done \& Page 2009).  We note that all of these ULXs have
luminosities $\la 3 \times 10^{39}$ erg s$^{-1}$, so can be close to
Eddington for moderately massive ($30-50 M_\odot$) stellar remnant
black holes, similar to that found in IC 10 X-1 (see introduction).
The next source in the sequence, NGC 1313 X-2, has a spectrum where
the high energy emission seems stronger than expected from a disc
dominated state, so this could instead be a type of very high state,
again with a moderately massive stellar remnant black hole. Hence, it
appears as though the low luminosity end of the ULX population could
potentially overlap with sub-Eddington processes seen in the BHB, albeit
for larger black holes.  We will explore this in a future paper by
characterising the properties of the Galactic BHBs at high Eddington
fractions in the {\it XMM-Newton\/} band pass.

The next category are those where there is clearly a soft inflection
as well as a high energy break, but where the total power still peaks
at the high energy break (Ho IX X-1; IC 342 X-1; NGC 1313 X-1).  These
are the ones where the simple disc plus (broken) power-law fits give a 
cool disc together with a high energy rollover, which we identify with
a new {\em ultraluminous} state. For these data the 
coupled disc-corona models give an intrinsic
disc temperature that is not given by either of these observed
characteristic energies, and where the effect of Comptonisation in an
optically thick, low temperature corona, is most marked.  These
spectra do not correspond to any of the known states, but can form
from a more extreme (higher optical depth, lower temperature) version
of the very high state corona.  These are typically brighter than
those in the previous class (though there is also substantial overlap
in luminosity) so are most likely super-Eddington accretion flows. 

There is then a clear observational sequence of spectral shapes,
through the sources where the ratio of power between Comptonisation
(the high energy peak) and the outer disc (low energy peak) steadily
decreases, from NGC 5204 X-1 and NGC 4559 X-1, to the most extreme
systems, namely Ho II X-1, NGC 5408 X-1 and NGC 55 ULX. These are the
ones where the inferred intrinsic disc temperature from the coupled
disc-corona model is $\la 0.4$~keV, far lower than expected from
stellar remnant accretion.

To understand these, we look first at what happens physically to the
flow as it approaches and then exceeds the Eddington limit.  It has
long been known that the Eddington limit for a disc is somewhat
different than that for spherical accretion (Shakura \& Sunyaev
1973). A thin disc, which radiates at the Eddington limit at all
radii, has an integrated luminosity $L \sim (1+ ln (\dot{m})) L_{\rm
Edd}$ where $\dot{m}$ = $\dot{M}/\dot{M}_{\rm Edd}$ is the mass
accretion rate scaled to that which gives the Eddington luminosity for
spherical accretion. However, this can only be achieved if excess
energy over and above that expected from a constant mass inflow rate,
radiating with constant efficiency, is somehow lost from the
system. There are two ways to do this, either by changing the mass
accretion rate as a function of radius through expelling the excess
mass via winds (Shakura \& Sunyaev 1973; Begelman et al. 2006) or by
changing the radiative efficiency by advecting the photons along with
the flow (slim discs, as above). Importantly, both can operate
simultaneously (Poutanen et al 2007), as indeed is shown in the most
recent 2D radiation hydrodynamic simulations of super-Eddington
accretion flows (Ohsuga 2006; 2007; 2009; Kawashima et al. 2009; Takeuchi , Mineshige, \& Ohsuga 2009).

This then gives a possible framework to interpret our results.  It is
clear that the objects with $\sim$~1~keV disc emission are probably
just more extreme versions of the brightest high and very high
spectral states known from BHBs.   The super-Eddington, ultraluminous state sources are then distinguished by their cool, optically thick coronae and apparently cooler discs.  One obvious source of higher optical
depth in the corona is the increasing importance of winds as the
source starts to accrete past the Eddington limit, and this mass
loading of the coronal particle acceleration mechanism leads to lower
temperatures of the Comptonising electrons.

These winds will become increasingly important as the flows become
increasingly super-Eddington, completely enveloping the inner regions
of the disc-corona out to an increasing photospheric radius (as in SS
433, Poutanen et al. 2007).  Hence the observed temperature decreases
in line with the Stefan-Boltzmann law (Shakura \& Sunyaev 1973;
Begelman et al. 2006; Poutanen et al. 2007). The outflow is inherently
(at least) two dimensional, so viewing angle will change the apparent
system luminosity (Ohsuga 2006; 2007).  We suggest the most extreme
objects seen in our current sample, NGC 4559 X-1, Ho II X-1, NGC 5408
X-1 and NGC 55 ULX\footnote{The relatively low luminosity of NGC 55
ULX may be an effect of its disc being close to edge-on to our
line-of-sight, evidence for which comes from the dipping behaviour
seen in its X-ray light curve (Stobbart et al. 2004).}, are dominated
by reprocessing in a wind, and that much of their luminosity output is
channeled into kinetic energy.

Thus it seems most likely that we are seeing the ULX transit between
the brightest high and very high states, through to a super-Eddington
{\em ultraluminous} state which is similar to the very high state but
with lower temperature and higher optical depth in the corona, to a
completely new (hyper-accreting) state where the wind dominates the
spectrum.  In none of these states do we require the presence of IMBHs
to explain the X-ray spectrum; all can be explained by stellar mass
black holes at high accretion rates, albeit perhaps black holes up to
a few times larger than those known in our own Galaxy.

\section{Conclusions}
\label{section:conclusion}

The highest quality data has been collated and utilised to both
characterise the spectra of ultraluminous X-ray sources and to
constrain their nature. These show that while some ULX (typically the
lowest luminosity ones) have spectra which are probably similar to the
high and (especially) the very high state in Galactic BHBs, the
majority show more complex curvature which can be modelled by a cool
disc component together with a power-law which breaks/rolls-over above
$\sim 3$ keV. This combination of spectral features is not commonly
present in any of the known (sub-Eddington) Galactic BHB states, and
we therefore propose these features as observational criteria for a
new {\it ultraluminous state} and identify it with super-Eddington
accretion flows.


More physical models for these spectra show they are not well fit by
(approximate, `$p$-free') slim disc models as these cannot
simultaneously produce both the soft excess and high energy break.
Instead, disc plus Comptonisation models give a much better
description of this complex curvature, indicating that the break above
$\sim 3$~keV comes from a cool, optically thick corona. This suggests
a more extreme version of the coronae seen in the very high state of
BHBs, as might be expected for super-Eddington flows.  However, such
coronae obscure the inner disc and alter its energetics (Done \&
Kubota 2006), so we model these effects to recover the intrinsic disc
temperatures.  Many of these are in the range expected for stellar
remnant black holes, showing that the apparent cool disc temperature
derived from simple disc models does not require an IMBH. However,
there are some objects where the recovered disc temperature (corrected
for the corona) is cooler than expected for a stellar remnant black
hole.  We suggest these are most likely to represent the most extreme
super-Eddington accretion flows, where the wind from the accretion
disc becomes so powerful that it envelops the inner disc out to a
large photospheric radius, producing the cool spectral component.

There are occasional spectra from BHBs at the highest luminosities
which are indeed better described by optically-thick Comptonisation
(GRO J1655-40 and GRS 1915+105: Makishima et al. 2000; Middleton et al.
2006; Ueda et al. 2009).  Similarly, the highest Eddington fraction AGN
such as RE J1034+396 (Middleton et al. 2009) and RX J0136-35 (Jin et
al. 2009) also show such spectra.  We suggest that all these sources
are in this new super-Eddington {\em ultraluminous}
accretion state.  It now appears
that ULXs, rather than revealing a new population of IMBHs, are
providing us with a template for accretion at super-Eddington rates.
This will have wide applications across many areas of astrophysics,
ranging from stellar formation to the growth of QSOs.  Further studies
of ULXs that provide a deeper understanding of this new and crucially
important accretion regime are therefore imperative.


\section*{Acknowledgements}

We thank the anonymous referee for their constructive comments, that have helped to improve this paper.  JCG gratefully acknowledges funding from the Science and Technology
Facilities Council (STFC) in the form of a PhD studentship. This work
is based on data from the {\it XMM-Newton} an ESA Science Mission with
instruments and contributions directly funded by ESA member states and
the USA (NASA).


\label{lastpage}

\end{document}